\documentclass[apj]{emulateapj}

\usepackage{graphicx}
\usepackage{t1enc}
\usepackage[varg]{txfonts}
\usepackage{epsfig}
\usepackage{rotating}
\usepackage{color}

\newcommand{\kms}     {km~s$^{-1}$}

\newcommand{\jpb}     {$\rm Jy~beam^{-1}$}    

\newcommand{\mo}      {M$_{\sun}$}

\newcommand{\chtoh}   {CH$_3$OH}
\newcommand{\water}   {H$_2$O}

\newcommand{\et}      {et al.}
\newcommand{\eg}      {e.\,g.,}
\newcommand{\ie}      {i.\,e.,}
\newcommand{\hii}     {\ion{H}{2}}
\newcommand{\uchii}   {UC\ion{H}{2}}

\newcommand{\phnn}    {\phantom{0}\phantom{0}}
\newcommand{\phnnn}   {\phantom{0}\phantom{0}\phantom{0}}

\newcommand{\phb}     {\phantom{$>$}}
\newcommand{\phv}     {\phantom{<}}

\newcommand{\raun}    {$^\mathrm{h~m~s}$}
\newcommand{\deun}    {$\mathrm{\degr~\arcmin~\arcsec}$}
\newcommand{\supa}    {$^\mathrm{a}$}
\newcommand{\supb}    {$^\mathrm{b}$}
\newcommand{\supc}    {$^\mathrm{c}$}

\shorttitle{Ionized gas in the massive star forming complex G75.78+0.34}
\shortauthors{S\'anchez-Monge et al.}

\begin{document}

\title{Deciphering the ionized gas content in the massive star forming complex
G75.78+0.34}

\author{\'Alvaro S\'anchez-Monge\altaffilmark{1,2,3},
        Stan Kurtz\altaffilmark{3},
        Aina Palau\altaffilmark{4},
        Robert Estalella\altaffilmark{2},
        Debra Shepherd\altaffilmark{5},
        Susana Lizano\altaffilmark{3},
        Jos\'e Franco\altaffilmark{6},
        Guido Garay\altaffilmark{7}}
\affil{$^{1}$ Osservatorio Astrofisico di Arcetri, INAF, Largo E.\ Fermi 5,
I-50125 Firenze, Italy; asanchez@arcetri.astro.it}
\affil{$^{2}$ Dpt d'Astronomia i Meteorologia (IEEC-UB), Institut de Ci\`encies
del Cosmos, Universitat de Barcelona, Mart\'i i Franqu\`es, 1, E-08028
Barcelona, Spain}
\affil{$^{3}$ Centro de Radioastronom\'ia y Astrof\'isica, Universidad
Nacional Aut\'onoma de M\'exico, Apdo. Postal 3-72, 58090, Morelia, Michoac\'an,
Mexico}
\affil{$^{4}$ Institut de Ci\`encies de l'Espai (CSIC-IEEC), Campus UAB --
Facultat de Ci\`encies, Torre C5p 2, E-08193 Bellaterra, Catalunya,
Spain}
\affil{$^{5}$ NRAO, P.O. Box O, Socorro, NM 87801-0387, USA}
\affil{$^{6}$ Instituto de Astronom\'ia, Universidad Nacional Aut\'onoma de
M\'exico, Apdo.~Postal 70-264, 04510 M\'exico, D.F., Mexico}
\affil{$^{7}$ Departamento de Astronom\'ia, Universidad de Chile, Camino el Observatorio 1515, Las Condes, Santiago, Chile}

\begin{abstract}
We present sub-arcsecond observations toward the massive star forming region G75.78+0.34. We used the Very Large Array to study the centimeter continuum and H$_2$O and CH$_3$OH maser emission, and the Owens Valley Radio Observatory and Submillimeter Array to study the millimeter continuum and recombination lines (H40$\alpha$ and H30$\alpha$). We found radio continuum emission at all wavelengths, coming from three components: (1) a cometary ultracompact (UC) \hii\ region with an electron density $\sim$$3.7\times10^4$~cm$^{-3}$, excited by a B0 type star, and with no associated dust emission; (2) an almost unresolved \uchii\ region (EAST), located $\sim$6\arcsec\ to the east of the cometary \uchii\ region, with an electron density $\sim$$1.3\times10^5$~cm$^{-3}$, and associated with a compact dust clump detected at millimeter and mid-infrared wavelengths; and (3) a compact source (CORE), located $\sim$2\arcsec\ to the southwest of the cometary arc, with a flux density increasing with frequency, and embedded in a dust condensation of 30~\mo. The CORE source is resolved into two compact and unresolved sources which can be well-fit by two homogeneous hypercompact \hii\ regions each one photo-ionized by a B0.5 ZAMS star, or by free-free radiation from shock-ionized gas resulting from the interaction of a jet/outflow system with the surrounding environment. The spatial distribution and kinematics of water masers close to the CORE-N and S sources, together with excess emission at 4.5~$\mu$m and the detected dust emission, suggest that the CORE source is a massive protostar driving a jet/outflow.
\end{abstract}


\section{Introduction} \label{sg75:intro}

Massive stars are key to understanding many physical phenomena in the Galaxy, however their first stages of formation are still poorly understood. One of the main reasons is that massive stars evolve more quickly to the main-sequence than low mass stars, and they radiate large amounts of ultraviolet (UV) photons and drive strong winds while they are still deeply embedded and accreting matter (\eg\ Beuther \& Shepherd 2005; Keto 2007). The interaction of the UV radiation and winds with the surrounding environment, resulting in bright radio continuum sources, must be well understood to comprehend the formation of high-mass stars.

Thermal (free-free) radio emission at centimeter wavelengths in regions of massive star formation can have distinct origins: {\it i)} \hii\ regions photoionized by embedded massive stars, with homogeneous density distributions (\eg\ Mezger \& Henderson 1967) or with density gradients (\eg\ Olnon 1975; Panagia \& Felli 1975; Franco \et\ 2000); {\it ii)} clumps of gas or circumstellar disks externally ionized by luminous early-type stars (\eg\ Garay 1987; O'Dell \& Wong 1996; Zapata \et\ 2004); {\it iii)} shock waves arising either in dense interstellar gas (\eg\ Ghavamian \& Hartigan 1998; Araya \et\ 2009) or from the collision of thermal radio jets from young stellar objects (YSOs) with their surroundings (\eg\ Anglada 1996; Eisloffel \et\ 2000; Rodr\'iguez \et\ 2005); {\it iv)} ionized accretion flows where the material becomes ionized while accreting onto the massive protostar (\eg\ Keto 2002, 2003, 2007); {\it v)} photoevaporated disks where the radiation of the newly-formed star ionizes and evaporates the surrounding disk (\eg\ Hollenbach \et\ 1994; Lugo \et\ 2004; \'Avalos \& Lizano 2012); {\it vi)} equatorial winds with the emission produced by small-scale ionized stellar winds (\eg\ Hoare 2006). All these mechanisms produce radio continuum sources with spectral indices, $\alpha$ ($S_\nu\propto\nu^\alpha$), between $-0.1$ and $+2$ (i.e., thermal emission; see Rodr\'iguez \et\ 1989). In addition, several works have found negative spectral indices (typical of non-thermal emission) associated with massive YSOs (\eg\ Rodr\'iguez \et\ 1989; Zapata \et\ 2006). These non-thermal sources can be young stars with active magnetospheres producing gyro-synchrotron emission (\eg\ Feigelson \& Montmerle 1999); synchrotron emission from fast shocks in disks or jets (\eg\ Reid \et\ 1995; Shepherd \& Kurtz 1999; Shchekinov \& Sobolev 2004); or extremely embedded YSOs where the UV photons from the massive protostars are heavily absorbed by large amounts of dust, with mass column densities $\gtrsim$10$^3$~g~cm$^{-2}$ (Rodr\'iguez \et\ 1993). A similar description of these mechanisms producing centimeter continuum emission can be found in Rodr\'iguez \et\ (2012).

It is possible --- and indeed, probable --- that several of these emission mechanisms, either thermal or non-thermal, occur within massive star formation regions,  either simultaneously or at different evolutionary epochs.  Any complete model of high-mass star formation must address the presence of these multiple modes of radio continuum emission.

\begin{table*}[ht!]
\vspace{-0.3cm}
\caption{Main continuum observational parameters of G75.78+0.34}
\centering
\begin{tabular}{c l l c c c c c c}
\hline\hline\noalign{\smallskip}
$\lambda$
&
&
&Epoch of
&Phase calibrator /
&Flux
&Beam
&P.A.
&rms
\\
(cm)
&Telescope
&Project
&Observation
&Bootstrapped flux\supa
&calibrator
&($\arcsec\times\arcsec$)
&($\degr$)
&(m\jpb)
\\
\hline
\noalign{\smallskip}
\phn6.0\phn	&VLA-B	&\texttt{AF381}		&2001 Apr 23		&2015+371 / $2.50(1)$\phn	&3C48		&$1.6\times1.3$		&$-87$		&0.49	\\
\phn3.6\phn	&VLA-B	&\texttt{AK440}		&1997 Mar 16		&2025+337 / $3.50(1)$\phn	&3C286		&$1.0\times0.7$		&\phn$+6$	&0.14	\\
\phn3.6\phn	&VLA-A	&\texttt{AK490}		&1999 Sep 12+13	&2015+371 / $2.01(1)$\phn	&3C286		&$0.3\times0.2$		&$-73$		&0.05	\\
\phn2.0\phn	&VLA-CnB	&\texttt{AK423}		&1996 Jan 29		&2025+337 / $2.72(2)$\phn	&3C286		&$1.7\times1.4$		&$+66$		&0.36	\\
\phn2.0\phn	&VLA-B	&\texttt{AF381}		&2001 Apr 23		&2015+371 / $2.62(1)$\phn	&3C286		&$0.6\times0.4$		&$-84$		&0.11	\\
\phn1.3\phn	&VLA-B	&\texttt{AK440}		&1997 Mar 16		&2025+337 / $3.48(2)$\phn	&3C286		&$0.31\times0.26$	&$-14$		&0.12	\\
\phn1.3\phn	&VLA-A	&\texttt{AK490}		&1999 Sep 12+13	&2015+371 / $1.82(2)$\phn	&3C286		&$0.10\times0.08$	&$+52$		&0.13	\\
\phn0.7\phn	&VLA-CnB	&\texttt{AK423}		&1996 Jan 29		&2025+337 / $2.33(2)$\phn	&3C286		&$2.5\times0.5$		&$-64$		&0.46	\\
\phn0.7\phn	&VLA-B	&\texttt{AK500}		&1999 + 2000		&2015+371 / $2.10(4)$\phn	&3C286		&$0.14\times0.12$	&$+37$		&0.25	\\
\phn0.31		&OVRO	&\texttt{}			&1997 + 1998		&BL~Lac / $5.4(20)$			&Uranus		&$2.0\times1.7$		&$-88$		&0.59	\\
\phn0.13		&OVRO	&\texttt{}			&1997 Sep--Dec	&BL~Lac / $2.5(30)$			&Uranus		&$1.6\times1.5$		&$-51$		&4.0\phn	\\
\phn0.12		&SMA		&\texttt{09B-S106}	&2010 Jun 10		&2025+337 / $1.7(15)$		&Callisto	&$6.4\times2.7$		&$-76$		&8.4\phn	\\
\hline
\end{tabular}
\begin{list}{}{}
\item[\supa] Bootstrapped flux density in Jy. The number in parentheses gives the uncertainty of the flux density as a percentage.
\end{list}
\label{tg75:obscont}
\end{table*}
\begin{table*}[ht!]
\vspace{-0.4cm}
\caption{Main observational parameters of the VLA and OVRO spectral line observations}
\centering
\begin{tabular}{l c c c c c c c c c c c}
\hline\hline\noalign{\smallskip}
&Frequency
&
&Bandwidth
&\multicolumn{2}{c}{Spectral resolution}
&$v_\mathrm{ref}$\supa
&Beam
&P.A.
&rms
\\
\cline{5-6}
Line
&(GHz)
&Telescope
&(MHz)
&(kHz)
&(\kms)
&(\kms)
&($\arcsec\times\arcsec$)
&($\degr$)
&(m\jpb)
\\
\hline\noalign{\smallskip}
\water\,(6$_{16}$-5$_{23}$)	&\phn22.235080	&VLA		&\phnnn1.56		&24.4		&0.33		&\phn$+1.$	&$0.07\times0.07$	&$-67$	&77.\phn	\\
							&				&VLA		&\phnnn6.25		&97.7		&1.3\phn		&$-10.$		&$0.10\times0.07$	&$-69$	&50.\phn	\\
\chtoh\,(7$_0$-6$_1$)		&\phn44.069410	&VLA		&\phnnn1.56		&24.4		&0.17		&\phn$+3.$	&$0.11\times0.10$	&$+32$	&30.\phn	\\
H40$\alpha$					&\phn99.022960	&OVRO	&\phn128.\phnn	&2000.\phnnn	&6.1\phn		&\phn$-8.$	&$5.6\times2.8$		&$+33$	&10.\phn	\\
H30$\alpha$					&231.900940		&OVRO	&\phn128.\phnn	&2000.\phnnn	&2.6\phn		&\phn$-8.$	&$3.9\times1.8$		&$+29$	&15.\phn	\\
\hline
\end{tabular}
\begin{list}{}{}
\item[\supa] Central velocity of the passband in the Local Standard of Rest.
\end{list}
\label{tg75:obsline}
\end{table*}

The ON-2 star forming complex contains several early-type (O and B) stars within a massive ($\ga 10^4$~\mo, Matthews \et\ 1986; Dent \et\ 1988) molecular cloud that spans $\sim$10\arcmin\ on the sky. Seen in CO, the cloud has two distinct condensations, with a roughly north-south orientation (Matthews \et\ 1986). Multiple ionized regions within the southern condensation were identified by Matthews \et\ (1973, 1977).  They reported an extended \hii\ region G75.77+0.34, and also the ultracompact (UC) \hii\ region G75.78$+$0.34 (hereafter G75); the latter is associated with strong OH maser emission first reported by Elld\'er \et\ (1969).

ON-2 lies toward Cygnus-X, and hence it has been problematic to assign a reliable distance.  Early estimates ranged from 0.9 to 5.5 kpc with the nearer distance coming from extinction or luminosity arguments and the farther distance coming from Galactic rotation models and radio recombination line velocities.  A helpful summary is presented by Campbell \et\ (1982).  More recent works, focusing on the UC \hii\ region rather than the molecular cloud, tend to adopt a kinematic distance of 5.6 kpc (\eg\ Wood \& Churchwell 1989; Hanson \et\ 2002). More recently, Ando \et\ (2011) observed the water masers associated with the G75 UC \hii\ region as part of the VERA (VLBI Exploration of Radio Astrometry) project.  They measured the trigonometric parallax and report a heliocentric distance of 3.83$\pm$0.13 kpc.  We consider this distance measurement to be the most reliable and we adopt 3.83 kpc for our analysis.  This distance places G75 close to other star forming regions (G75.76+0.35 and AFGL\,2591; Rygl \et\ 2012), and close to the solar circle (Ando \et\ 2011; Nagayama \et\ 2012).

At centimeter wavelengths, G75 is dominated by a cometary \uchii\ region reported by Wood \& Churchwell (1989). Hofner \& Churchwell (1996) detected a cluster of water masers located about 2\arcsec\ ($10^4$~AU) southwest of the \uchii\ region, coincident with a compact radio continuum source (Carral \et\ 1997) with a spectral index of $+1.5\pm0.4$ from 6~cm through 3~mm. Franco \et\ (2000) modeled this compact continuum source as a hypercompact (HC) \hii\ region with $n_{e}\propto r^{-4}$.  They note that this very steep density gradient is probably unrealistic, and mentioned several possible causes, including a contribution from warm dust to the flux density at 0.7~cm. Additionally, emission from a myriad of molecular line transitions has been reported in single-dish surveys (\eg\ Shirley \et\ 2003, Roberts \& Millar 2007, Klaassen \& Wilson 2007, Bisschop \et\ 2007). Higher angular resolution observations of different dense gas tracers show that most of the molecular emission comes from the compact radio continuum source associated with the cluster of water masers (Codella \et\ 2010; S\'anchez-Monge 2011). Finally, up to four distinct outflows have been identified in the ON-2 cloud core (Shepherd \et\ 1997). All this makes G75 an excellent target to study the nature of the centimeter continuum sources in a massive star forming region.

In this paper, we present high angular resolution centimeter continuum observations together with 22~GHz \water\ and 44~GHz \chtoh\ maser observations toward G75. We complement this data with millimeter continuum and radio recombination line observations, with the goal of deciphering the nature of the centimeter continuum sources.

\section{Observations} \label{sg75:obs}

\subsection{VLA radio continuum observations} \label{sg75:vlacont}

G75.78+0.34 was observed with the Very Large Array (VLA\footnote{The Very Large Array (VLA) is operated by the National Radio Astronomy Observatory (NRAO), a facility of the National Science Foundation operated under cooperative agreement by Associated Universities, Inc.}) at 6.0, 3.6, 2.0, 1.3, and 0.7~cm from January 1996 to April 2001, using the array in the CnB, B, and A configurations. In Table~\ref{tg75:obscont} we summarize these observations. The data reduction followed  standard procedures for calibration of high frequency data, using the NRAO package AIPS. Initial images were produced with a robust parameter (Briggs 1995) of 1 (see Table~\ref{tg75:obscont} for synthesized beams and rms noise levels of these images). These data can be grouped into two categories based on angular resolution: those with $\theta_\mathrm{beam}\geq\!1\farcs0$, and those with $\theta_\mathrm{beam}\sim$$0\farcs2$. The continuum images at 1.3~cm and 0.7~cm, from projects AK490 and AK500, were cross-calibrated with the strongest \water\ and \chtoh\ maser components, respectively, which were observed simultaneously with the continuum emission (see Section~\ref{sg75:vlaline} and Table~\ref{tg75:obsline}). After comparing the initial maps for consistency, we combined the \emph{uv}--data at the same frequencies to obtain final images with better \emph{uv}-coverage and sensitivity. The resulting synthesized beams and rms noise levels of the combined images are given in Table~\ref{tg75:rescont}. 

At 6~cm, the \hii\ region G75.77+0.34, approximately $1'$ to the southwest (e.g., Riffel \& L\"udke 2010), produced non-imageable artifacts. The shortest baselines ($<\!5$~k$\lambda$) were excluded, producing essentially no change in the measured flux density but significantly improving the quality of the map.

\subsection{VLA \water\ and \chtoh\ maser observations} \label{sg75:vlaline}

The water maser line at 22.23508~GHz ($6_{16}-5_{23}$ transition) was observed with the VLA in the A configuration (project AK490) simultaneously with the 1.3~cm continuum emission. Two different correlator configurations were used, providing spectral resolutions of 0.3 and 1.3~\kms; with velocity coverages of 21 and 83~\kms, respectively. 

The class~I methanol maser line at 44.06941~GHz ($7_{0}-6_{1}$ A$^{+}$ transition) was observed with the VLA in the B configuration (project AK500) simultaneously with the 0.7~cm continuum emission. Details of the spectrometer configuration are given in Table~\ref{tg75:obsline}. 

The \water\ and \chtoh\ maser data were calibrated following the AIPS guidelines for calibration of high frequency data. Self-calibration was performed on the strongest maser component, and the solutions were applied to the spectral line and continuum data. The images were constructed using uniform and natural weighting to measure the maser positions at the highest angular resolution and to estimate the intensity of the different maser components, respectively.

\begin{figure}[t]
\begin{center}
\begin{tabular}[b]{c}
\vspace{0.1cm}
       \epsfig{file=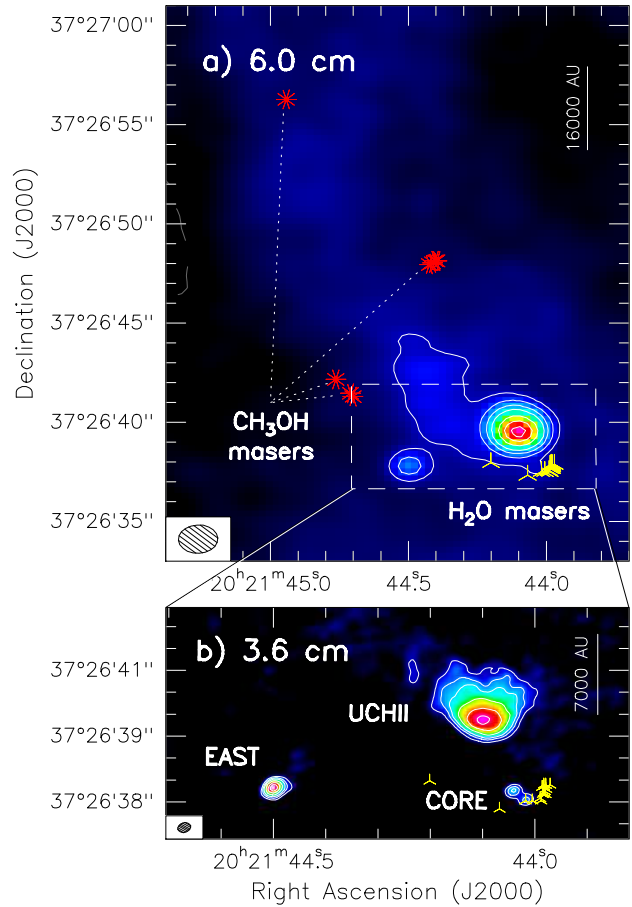, scale=0.88} \\
\end{tabular}
\caption{G75.78+0.34 star-forming region. {\bf (a)}: VLA 6.0~cm continuum image. Levels are $-3$, 3, 6, 10, 15, 25, 35, 45, and 55 times 490~$\mu$\jpb. Ten-point red stars indicate \chtoh\ masers, and three-point yellow stars indicate \water\ masers. The dashed box shows the region zoomed in the bottom panel. {\bf (b)}: VLA 3.6~cm continuum image. Levels are $-4$, 4, 8, 12, 20, 30, and 40 times 50~$\mu$\jpb. Synthesized beams of the continuum images are shown in the bottom-left corner; spatial scales are indicated in the top-right corner of each panel.}
\label{fg75:region}
\end{center}
\end{figure}

\begin{figure*}[htpb!]
\begin{center}
\begin{tabular}[b]{c}
\vspace{0.1cm}
       \epsfig{file=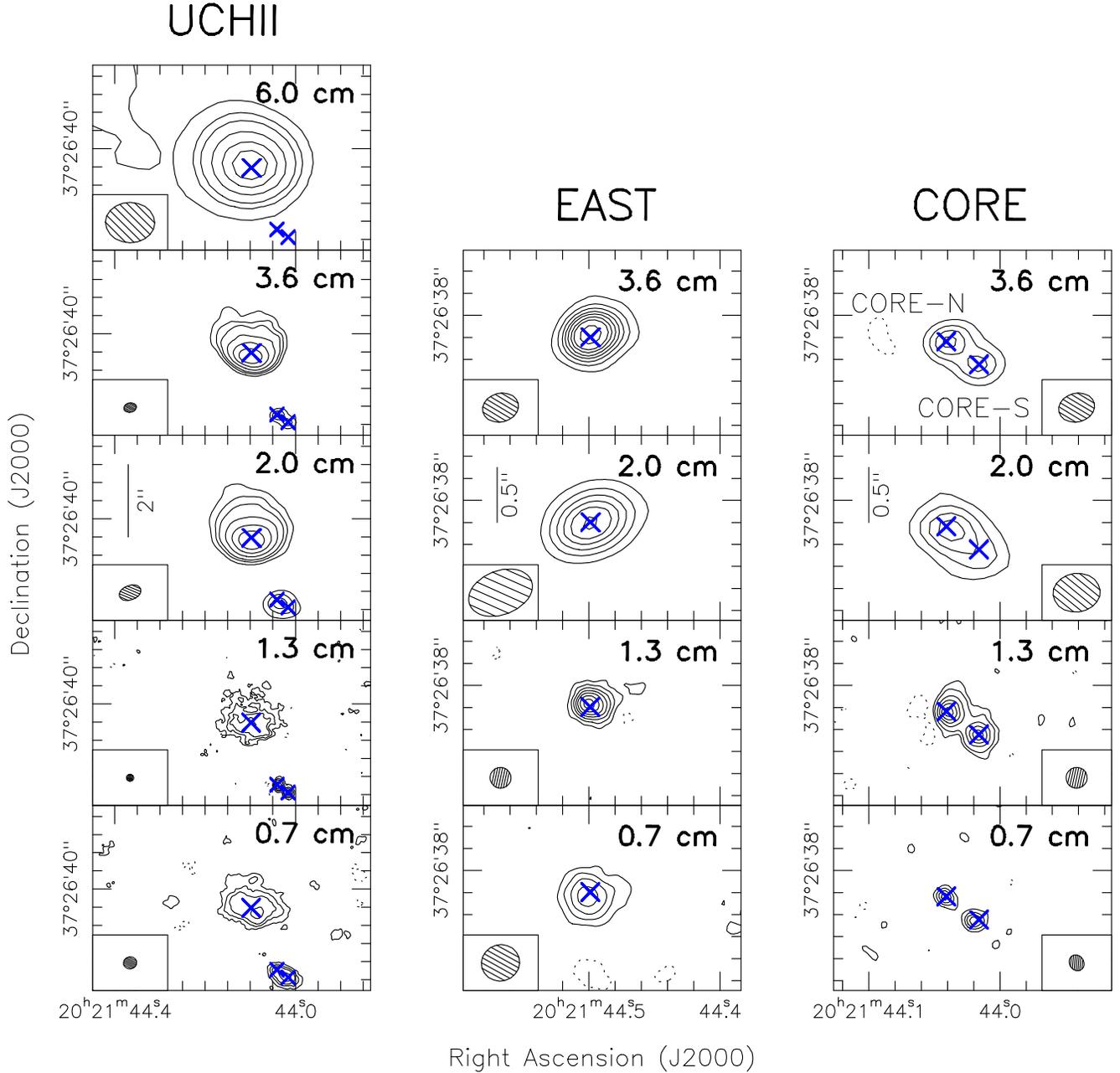, scale=0.98} \\
\end{tabular}
\caption{
Close-up continuum images of the UCHII, EAST and CORE sources. In each panel we show the VLA image at the wavelength indicated in the top-right corners. {\bf Left column}: UCHII source. For all panels, levels are $-4$, 4, 8, 12, 20, 30, and 40 times the rms noise level of the map: 485, 61, 138, 67, and 159~$\mu$\jpb, for 6.0, 3.6, 2.0, 1.3, and 0.7~cm maps, respectively. {\bf Center column}: EAST source. For all panels, levels are $-4$, 4, 8, 12, 16, 20, 25, and 30 times the rms noise level of the map: 61, 138, 67, and 159~$\mu$\jpb, for 3.6, 2.0, 1.3, and 0.7~cm maps, respectively. {\bf Right column}: CORE (N and S) source. For all panels, levels are $-3$, 3, 6, 12, 18, 24, and 30 times the rms noise level of the map: 61, 144, 67, and 186~$\mu$\jpb, for 3.6, 2.0, 1.3, and 0.7~cm maps, respectively. See Table~\ref{tg75:rescont} for details of the synthesized beams, and the flux densities and sizes of the sources. Blue crosses in all panels indicate the position of the four continuum sources: UCHII, EAST, CORE-N and CORE-S (see Table~\ref{tg75:rescont}). The scale in arcseconds is shown in the 2.0~cm panels. At the adopted distance of 3.8~kpc, $2''$ corresponds to 7600 AU while $0.5''$ is 1900 AU.}
\label{fg75:cont}
\end{center}
\end{figure*}

\subsection{OVRO observations} \label{sg75:ovro}

The Owens Valley Radio Observatory (OVRO\footnote{OVRO was operated by the California Institute of Technology with support from the National Science Foundation.}) observations at 3 and 1~mm were made in the L (Low) and H (High) resolution configurations during September, October and December 1997. In March 1998, additional 3~mm observations were made in the uH (ultra-High) resolution configuration. All the observations were made in the double sideband mode, simultaneously observing at 3 and 1~mm. The continuum was observed in two 1~GHz channels, centered at 95.78~GHz and 98.78~GHz for 3~mm and 228.85~GHz and 231.85~GHz for 1~mm. 

In addition, spectral line modules covered the H40$\alpha$ (99.02296~GHz) and H30$\alpha$ (231.90094~GHz) radio recombination lines (RRLs). Each spectral line setup consisted of 62 Hanning smoothed channels of 2~MHz each, providing resolutions of 6.1 and 2.6~\kms\ at 3~mm and 1~mm, respectively. The assumed LSR velocity for both lines was $-8$~\kms.  Bandpass calibration was performed by observing the quasar 3C454.3. Amplitude and phase calibration were achieved by monitoring BL~Lac during the different observing tracks. The absolute flux density scale was determined from Uranus, with an estimated uncertainty of 20\% at 3~mm and 30\% at 1~mm. The data were reduced using a combination of routines from the OVRO reduction package MMA, from MIRIAD, and from AIPS. Imaging was performed with the task IMAGR of AIPS. In Tables~\ref{tg75:obscont} and \ref{tg75:obsline} we list details of the continuum and radio recombination line observations.

\subsection{SMA observations} \label{sg75:sma}

G75.78+0.34 was observed with the Submillimeter Array (SMA\footnote{The SMA is a joint project between the Smithsonian Astrophysical Observatory and the Academia Sinica Institute of Astronomy and Astrophysics, and is funded by the Smithsonian Institution and the Academia Sinica.}) in the 1.3~mm (230~GHz) band using the compact configuration on 2010 June 10. A total bandwidth of $2\times4$~GHz was used, covering the frequency ranges 218.2--222.2~GHz and 230.2--234.3~GHz, with a spectral resolution of $\sim$1~\kms. System temperatures ranged between 150 and 250~K. The zenith opacities at 225~GHz were around 0.10 and 0.15 during the 3-hour track. The FWHM of the primary beam at 1.3~mm was $\sim$56\arcsec. Bandpass calibration was performed by observing the quasar 3C454.3. Amplitude and phase calibrations were made by monitoring 2025+337 and 2015+371, with an rms phase of $\sim$40\degr. The absolute flux density scale was determined from Callisto with an estimated uncertainty around 15\%. Data were calibrated and imaged with the MIRIAD software package. The continuum was constructed in the ({\it u,v}) domain from the line-free channels. Imaging was performed using natural weighting, resulting in a synthesized beam of $6\farcs4\times2\farcs7$ with a P.A.=$-76$\degr, and 1~$\sigma$ rms of 8.4~mJy~beam$^{-1}$ for the continuum. The two 4 GHz passbands include several molecular transitions, including CO, CH$_3$CN, and CH$_3$CCH. These molecular line data  will be presented in a forthcoming paper together with VLA ammonia observations (S\'anchez-Monge \et, in prep.).

\begin{table*}[htpb!]
\caption{Multiwavelength results for the YSOs in the star-forming region G75.78+0.34}
\centering
\begin{tabular}{c c c c c c c c c c}
\hline\hline\noalign{\smallskip}	
Wavelength
&Beam
&P.A.
&Rms
&$I_\nu^\mathrm{peak}$\supa
&$S_\nu$\supa
&Deconv. Size
&P.A.
\\
(cm)
&($\arcsec\times\arcsec$)
&($\degr$)
&($\mu$\jpb)
&(mJy~beam$^{-1}$)
&(mJy)
&($\arcsec\times\arcsec$)
&($\degr$)
\\
\hline\hline
\noalign{\smallskip}
\multicolumn{1}{l}{\uchii}	&\multicolumn{3}{l}{$\alpha(J2000.0)=20^\mathrm{h}21^\mathrm{m}44\fs098$\supb}	\\
				&\multicolumn{3}{l}{$\delta(J2000.0)=+37\degr26\arcmin39\farcs47$\supb}		\\
\hline
\noalign{\smallskip}
6.0\phn		&$1.30\times1.11$	&$+89$		&\phn$485$		&$24\pm1$		&$49\pm6$\phn	&$1.3\times1.2\phn\phv~\pm0.4$	&\phn$60\pm40$	\\
3.6\phn		&$0.33\times0.26$	&$-75$		&\phnn$61$		&$\phn3\pm1$		&$32\pm4$\phn	&$1.0\times0.8\phn\phv~\pm0.1$	&\phn$70\pm10$	\\
2.0\phn		&$0.60\times0.40$	&$-69$		&\phn$138$		&$\phn7\pm1$		&$34\pm4$\phn	&$1.1\times0.9\phn\phv~\pm0.1$	&\phn$70\pm10$	\\
1.3\phn		&$0.19\times0.19$	&$+33$		&\phnn$67$		&$\phn2\pm1$		&$32\pm4$\phn	&$1.0\times0.9\phn\phv~\pm0.1$	&\phn$80\pm10$	\\
0.7\phn		&$0.35\times0.33$	&$-74$		&\phn$159$		&$\phn4\pm1$		&$27\pm4$\phn	&$1.2\times0.8\phn\phv~\pm0.2$	&\phn$75\pm10$	\\
0.31			&$1.96\times1.74$	&$-88$		&\phn$579$		&$21\pm2$		&$34\pm14$		&$2.0\times0.8\phn\phv~\pm0.3$	&$120\pm10$		\\
0.13			&$1.61\times1.49$	&$-51$		&$4000$			&\ldots			&$<16$			&\ldots							&\ldots			\\
\hline\hline
\noalign{\smallskip}
\multicolumn{1}{l}{EAST}	&\multicolumn{3}{l}{$\alpha(J2000.0)=20^\mathrm{h}21^\mathrm{m}44\fs499$\supb}	\\
				&\multicolumn{3}{l}{$\delta(J2000.0)=+37\degr26\arcmin37\farcs80$\supb}		\\
\hline
\noalign{\smallskip}
6.0\phn		&$1.30\times1.11$	&$+89$			&\phn$485$		&$\phn4.1\pm1.0$	&$\phn4.5\pm1.5$	&$1.3\times1.0\phn\phv~\pm0.7$	&$130\pm50$		\\
3.6\phn		&$0.33\times0.26$	&$-76$			&\phnn$61$		&$\phn2.9\pm0.2$	&$\phn4.5\pm0.6$	&$0.25\times0.20\phv~\pm0.04$	&$135\pm40$		\\
2.0\phn		&$0.60\times0.40$	&$-69$			&\phn$138$		&$\phn3.8\pm0.3$	&$\phn3.9\pm0.7$	&$0.15\times<0.3\phn~\pm0.15$	&$105\pm40$		\\
1.3\phn		&$0.19\times0.19$	&$+33$			&\phnn$67$		&$\phn2.2\pm0.2$	&$\phn3.9\pm0.6$	&$0.20\times0.15\phv~\pm0.05$	&\phn$90\pm20$	\\
0.7\phn		&$0.35\times0.33$	&$-74$			&\phn$159$		&$\phn2.9\pm0.4$	&$\phn3.3\pm0.7$	&$0.2\times0.1\phn\phv~\pm0.2$	&\phn$90\pm50$	\\
0.31			&$1.96\times1.74$	&$-88$			&\phn$579$		&$\phn6.0\pm1.2$	&$\phn9.9\pm4.7$	&$2.8\times1.7\phn\phv~\pm0.5$	&\phn$25\pm20$	\\
0.13			&$1.61\times1.49$	&$-51$			&$4000$			&\ldots			&$<17$			&\ldots							&\ldots			\\
\hline\hline
\noalign{\smallskip}
\multicolumn{1}{l}{CORE}	&\multicolumn{3}{l}{$\alpha(J2000.0)=20^\mathrm{h}21^\mathrm{m}44\fs030$\supb}	\\
				&\multicolumn{3}{l}{$\delta(J2000.0)=+37\degr26\arcmin37\farcs67$\supb}		\\
\hline
\noalign{\smallskip}
6.0\phn		&$1.30\times1.11$	&$+89$			&\phn$485$		&\ldots			&$<1.9$			&\ldots							&\ldots			\\
3.6\phn		&$0.33\times0.26$	&$-76$			&\phnn$61$		&$\phn0.8\pm0.2$	&$\phn1.4\pm0.3$	&\ldots							&\ldots			\\
2.0\phn		&$0.60\times0.40$	&$-69$			&\phn$138$		&$\phn2.1\pm0.3$	&$\phn3.1\pm0.7$	&$0.45\times0.23\phv~\pm0.09$	&\phn$50\pm20$	\\
1.3\phn		&$0.19\times0.19$	&$+33$			&\phnn$67$		&$\phn2.2\pm0.2$	&$\phn4.8\pm0.7$	&\ldots							&\ldots			\\
0.7\phn		&$0.35\times0.33$	&$-74$			&\phn$159$		&$\phn4.0\pm0.3$	&$\phn7.9\pm1.2$	&$0.6\times0.2\phn\phv~\pm0.1$	&\phn$55\pm15$	\\
0.31			&$1.96\times1.74$	&$-88$			&\phn$579$		&$23\pm2$		&$\phn36\pm14$	&$2.0\times1.4\phn\phv~\pm0.3$	&$145\pm10$		\\
0.13			&$1.61\times1.49$	&$-51$			&$4000$			&$\phn98\pm10$	&$\phn300\pm100$	&$4.0\times2.2\phn\phv~\pm0.3$	&$145\pm10$		\\
\hline\hline
\noalign{\smallskip}
\multicolumn{1}{l}{CORE-N}	&\multicolumn{3}{l}{$\alpha(J2000.0)=20^\mathrm{h}21^\mathrm{m}44\fs041$\supb}	\\
				&\multicolumn{3}{l}{$\delta(J2000.0)=+37\degr26\arcmin37\farcs76$\supb}		\\
\hline
\noalign{\smallskip}
3.6\phn		&$0.33\times0.26$	&$-76$			&\phnn$61$		&$\phn0.8\pm0.2$	&$\phn0.8\pm0.2$	&$0.4\times0.1\phn\phv~\pm0.1$	&\phn$50\pm20$	\\
2.0\phn	 	&$0.43\times0.35$	&$-84$			&\phn$144$		&$\phn2.0\pm0.3$	&$\phn2.1\pm0.5$	&$0.4\times0.1\phn\phv~\pm0.1$	&\phn$50\pm15$	\\
1.3\phn		&$0.19\times0.19$	&$+33$			&\phnn$67$		&$\phn2.2\pm0.2$	&$\phn2.7\pm0.4$	&$0.15\times0.08\phv~\pm0.04$	&\phn$30\pm10$	\\
0.7\phn		&$0.15\times0.13$	&$+40$			&\phn$186$		&$\phn2.6\pm0.4$	&$\phn2.6\pm0.8$	&$0.05\times<0.07~\pm0.05$	&\phn$55\pm30$	\\
\hline\hline
\noalign{\smallskip}
\multicolumn{1}{l}{CORE-S}	&\multicolumn{3}{l}{$\alpha(J2000.0)=20^\mathrm{h}21^\mathrm{m}44\fs016$\supb}	\\
				&\multicolumn{3}{l}{$\delta(J2000.0)=+37\degr26\arcmin37\farcs55$\supb}		\\
\hline
\noalign{\smallskip}
3.6\phn		&$0.33\times0.26$	&$-76$			&\phnn$61$		&$\phn0.6\pm0.2$	&$\phn0.6\pm0.2$	&$0.5\times0.1\phn\phv~\pm0.1$	&\phn$55\pm15$	\\
2.0\phn	 	&$0.43\times0.35$	&$-84$			&\phn$144$		&$\phn1.8\pm0.3$	&$\phn1.8\pm0.5$	&$0.4\times0.1\phn\phv~\pm0.1$	&\phn$50\pm15$	\\
1.3\phn		&$0.19\times0.19$	&$+33$			&\phnn$67$		&$\phn1.9\pm0.2$	&$\phn2.1\pm0.4$	&$0.20\times0.07\phv~\pm0.05$	&\phn$45\pm10$	\\
0.7\phn		&$0.15\times0.13$	&$+40$			&\phn$186$		&$\phn2.7\pm0.4$	&$\phn3.2\pm0.9$	&$0.10\times<0.07~\pm0.05$		&$100\pm30$		\\
\hline
\end{tabular}
\begin{list}{}{}
\small
\item[\supa] Primary beam corrected. Error in intensity is $2\sigma$. Error in flux density has been calculated as $\big[\big(2\sigma\left(\theta_\mathrm{source}/\theta_\mathrm{beam}\right)^{1/2}\big)^{2}+\big(2\sigma_\mathrm{flux-scale}\big)^{2}\big]^{1/2}$, where $\sigma$ is the rms noise level of the map, $\theta_\mathrm{source}$ and $\theta_\mathrm{beam}$ are the size of the source and the beam respectively, and $\sigma_\mathrm{flux-scale}$ is the error in the flux scale, which takes into account the uncertainty on the calibration applied to the flux density of the source ($S_\nu\times\%_\mathrm{uncertainty}$).
\item[\supb] Coordinates calculated as an average of the coordinates at each wavelength where the source is well detected (\ie\ not filtered out). Typical discrepancies in the coordinates at different wavelengths are $<0\farcs1$.
\end{list}
\label{tg75:rescont}
\end{table*}

\begin{figure*}[t!]
\begin{center}
\begin{tabular}[b]{c}
       \epsfig{file=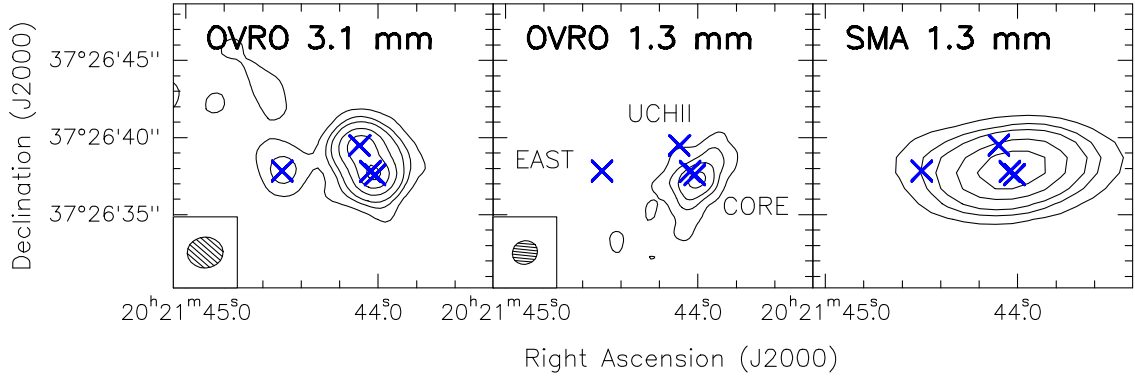, scale=1.09} \\
\end{tabular}
\caption{OVRO and SMA millimeter continuum maps. Levels are $-4$, 4, 8, 12, 20, 30, and 40 times the rms noise level of the map: 0.6, 4, and 8.4~m\jpb, for OVRO 3.1~mm, OVRO 1.3~mm, and SMA 1.3~mm maps, respectively. See Table~\ref{tg75:rescont} and Section~\ref{sg75:resmm} for details of the flux and beam of each image. Blue crosses show the position of the radiocontinuum sources, as in Figure~\ref{fg75:cont}.}
\label{fg75:contmm}
\end{center}
\end{figure*}
\begin{figure*}[t!]
\begin{center}
\begin{tabular}[b]{c}
       \epsfig{file=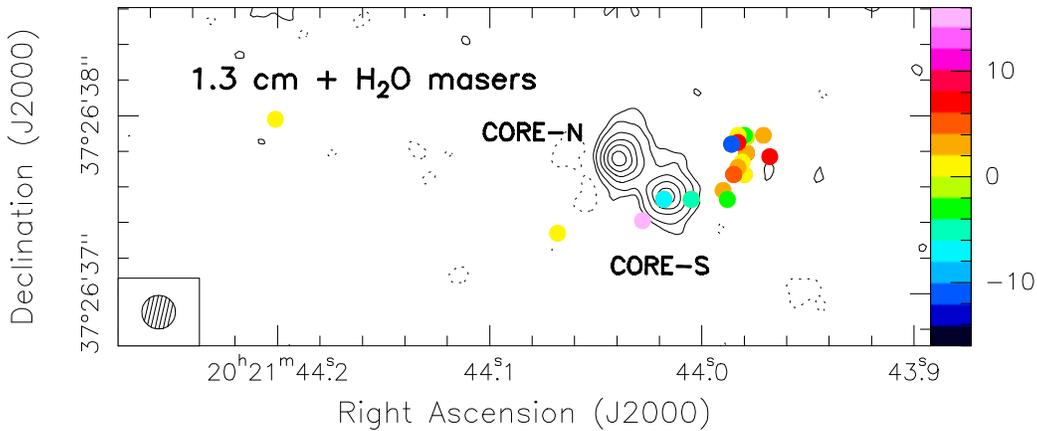, scale=0.9} \\
\end{tabular}
\caption{\water\ maser distribution in G75. {\bf Black} contours: VLA 1.3~cm continuum image as in Figure~\ref{fg75:cont} -- middle column. Color circles show the position of the 35 water maser spots (see Table~\ref{tg75:masers}). Different colors are used to indicate the maser LSR velocities, according to the color scale (in \kms) on the right hand side of the plot.}
\label{fg75:masers}
\end{center}
\end{figure*}

\section{Results} \label{sg75:res}

\subsection{Centimeter continuum emission} \label{sg75:rescm}

We detected centimeter radio continuum emission at all wavelengths. In Figure~\ref{fg75:region} we show a global overview of the G75 region at 6.0~cm (panel a) and 3.6~cm (panel b). As seen in the figure, the centimeter continuum emission is dominated by three components that we name UCHII, EAST and CORE. In Figure~\ref{fg75:cont} we show a close-up of these three sources for the combined maps at each frequency, and in Table~\ref{tg75:rescont} we list their flux densities and sizes, together with the beams, and rms noise levels of the combined images. 

The strongest source (UCHII; G75.7826$+$0.3429) is a cometary \uchii\ region with an integrated flux density of $\sim$35~mJy and an angular diameter of $\sim$1\arcsec\ ($\sim$0.02~pc at a distance of 3.8~kpc; previously imaged by Wood \& Churchwell 1989 and Carral \et\ 1997). Located $\sim$6\arcsec\ to the east, we identify a compact source (EAST; G75.7830$+$0.3416), with a flux density of $\sim$4~mJy and an angular diameter of $\sim$$0\farcs2$ ($\sim$760~AU at a distance of 3.8~kpc). This source appears at the $\sim$5$\sigma$ level in the 7~mm maps of Carral \et\ (1997). Finally, at the head of the cometary arc we find a compact source (CORE; G75.7821$+$0.3418) slightly elongated in the northeast-southwest direction, with a flux density of a few mJy and increasing with frequency. This source is coincident with the 7~mm continuum source reported by Carral \et\ (1997) and with the clump of \water\ masers (Hofner \& Churchwell 1996). Our higher angular resolution observations ($\le$$0\farcs3$) allow us to resolve the CORE source into two distinct (marginally resolved) condensations: CORE-N (G75.7821$+$0.3428) and CORE-S (G75.7820$+$0.3429; see Figure~\ref{fg75:cont}).

\subsection{Millimeter continuum emission} \label{sg75:resmm}

In Figure~\ref{fg75:contmm} we show the OVRO and SMA millimeter continuum images of the G75 region. At 3.1~mm we detect emission from all three sources, with the emission from the UCHII and CORE sources barely resolved. At 1.3~mm we only detect emission associated with the CORE source, probably due to a lack of sensitivity that precludes the detection of faint emission from the EAST and UCHII sources. The angular resolution of our millimeter continuum images (10--15 times poorer than the angular resolution at centimeter wavelengths) does not allow us to resolve the components CORE-N and CORE-S. In Table~\ref{tg75:rescont} we list the flux densities and sizes, and the beams and rms noise levels of the OVRO observations. For the SMA 1.3~mm source, we fit a two-dimensional Gaussian obtaining a primary beam corrected flux density of $0.60\pm0.17$~Jy with a deconvolved size of $(4\times2\pm1)$\arcsec\ at a P.A.=$(75\pm10)\degr$. The SMA 1.3~mm source, although centered on the CORE position, encompasses the UCHII source and shows an extension to the EAST source. In addition, the flux density measured in the SMA image is almost twice the flux density measured in the OVRO image, suggesting that there is some faint, extended emission not detected in the OVRO map. We refrain from combining the OVRO and SMA 1.3~mm continuum images because of the poorer {\it uv}--coverage and sensitivity of the latter. Higher sensitivity maps are needed to confirm the 1~mm continuum emission associated with the UCHII and EAST sources, and higher angular resolution is needed to identify the millimeter emission of CORE-N and S sources.

At 3.5~mm Riffel \& L\"udke (2010) measured a flux density of 119~mJy with a synthesized beam of 18\arcsec, while Shepherd \et\ (1997) measured 75~mJy with a resolution of 5\arcsec, both observations made with the BIMA telescope. Our measurement ($\sim$75~mJy, including all the flux at 3.1~mm) is in good agreement with that of Shepherd \et\, but somewhat lower than that of Riffel \& L\"udke.  This difference may arise because the latter work (with 18\arcsec resolution) was more sensitive to weak, extended emission.

\begin{table}[t!]
\caption{22~GHz water and 44~GHz methanol masers in G75.78+0.34}
\centering
\footnotesize
\begin{tabular}{c c c c c c}
\hline\hline\noalign{\smallskip}
$\alpha$(J2000.0)
&$\delta$(J2000.0)
&$S_\mathrm{peak}$
&$v_\mathrm{peak}$
&$\int S \mathrm{d}v$
&
\\
(~\raun~)
&(~\deun~)
&(Jy)
&(\kms)
&(Jy \kms)
&Notes$^\mathrm{a}$
\\
\hline\hline
\noalign{\smallskip}
\multicolumn{5}{l}{\phn \water\ maser components}   \\
\hline
\noalign{\smallskip}
20 21 43.968   &37 26 37.77    &\phnn0.3\phn   &\phn$+7.0$     &\phnn0.4\phn   &L      \\
20 21 43.971   &37 26 37.89    &\phnn1.5\phn   &\phn$+3.0$     &\phnn1.7\phn   &H      \\
20 21 43.979   &37 26 37.79    &\phnn3.9\phn   &\phn$+6.9$     &\phnn3.1\phn   &H      \\
               &               &\phn15.0\phn   &\phn$+2.3$     &\phn16.4\phn   &H,L    \\
20 21 43.979   &37 26 37.88    &\phnn2.4\phn   &\phn$+0.0$     &\phnn1.4\phn   &H      \\
               &               &\phnn1.0\phn   &\phn$-0.8$     &\phnn1.3\phn   &L      \\
20 21 43.980   &37 26 37.67    &\phnn4.5\phn   &\phn$+2.6$     &\phnn6.3\phn   &H      \\
               &               &\phnn0.2\phn   &\phn$+0.5$     &\phnn0.3\phn   &L      \\
20 21 43.980   &37 26 37.89    &\phnn1.5\phn   &\phn$+0.5$     &\phnn2.0\phn   &L      \\
               &               &\phnn0.7\phn   &\phn$-2.6$     &\phnn0.6\phn   &H,L    \\
20 21 43.990   &37 26 37.58    &\phnn7.5\phn   &\phn$+3.0$     &\phn10.5\phn   &H,L    \\
20 21 43.981   &37 26 37.74    &\phnn5.1\phn   &\phn$+1.7$     &\phnn2.6\phn   &H      \\
20 21 43.983   &37 26 37.71    &122.0\phn      &\phn$+3.0$     &220.0\phn      &H      \\
20 21 43.983   &37 26 37.89    &\phnn1.7\phn   &\phn$+1.3$     &\phnn0.9\phn   &H      \\
20 21 43.983   &37 26 37.85    &\phn53.0\phn   &\phn$+6.9$     &111.0\phn      &H,L    \\
20 21 43.985   &37 26 37.67    &\phnn0.2\phn   &\phn$+6.9$     &\phnn0.12      &H      \\
               &               &\phnn0.3\phn   &\phn$+4.6$     &\phnn0.6\phn   &H,L    \\
20 21 43.986   &37 26 37.84    &\phnn6.5\phn   &\phn$+4.3$     &\phnn3.9\phn   &H,L    \\
               &               &\phnn0.6\phn   &$-11.0$        &\phnn1.6\phn   &L      \\
20 21 43.988   &37 26 37.53    &\phnn3.1\phn   &\phn$+0.3$     &\phnn2.5\phn   &H,L    \\
               &               &\phnn1.0\phn   &\phn$-2.3$     &\phnn1.0\phn   &H      \\
20 21 44.005   &37 26 37.53    &\phnn0.6\phn   &\phn$+5.5$     &\phnn0.6\phn   &H,L    \\
               &               &\phnn1.1\phn   &\phn$+1.7$     &\phnn0.9\phn   &H      \\
               &               &\phnn7.2\phn   &\phn$-1.3$     &\phn15.1\phn   &H,L    \\
               &               &\phn15.0\phn   &\phn$-5.6$     &\phn20.4\phn   &H,L    \\
20 21 44.017   &37 26 37.58    &\phnn0.5\phn   &$-30.0$        &\phnn0.7\phn   &L      \\
               &               &\phnn1.4\phn   &$-36.0$        &\phnn1.8\phn   &L      \\
20 21 44.018   &37 26 37.53    &\phn47.0\phn   &\phn$+7.3$     &\phn65.5\phn   &H,L    \\
               &               &\phnn0.4\phn   &\phn$+4.9$     &\phnn0.2\phn   &H      \\
               &               &\phnn2.7\phn   &\phn$+2.0$     &\phnn3.0\phn   &H      \\
               &               &\phn56.0\phn   &\phn$-0.7$     &150.0\phn      &H,L    \\
               &               &\phnn3.7\phn   &\phn$-6.2$     &\phnn6.7\phn   &H,L    \\
20 21 44.028   &37 26 37.41    &\phnn0.7\phn   &$+14.0$        &\phnn0.9\phn   &L      \\
20 21 44.068   &37 26 37.34    &\phnn2.3\phn   &\phn$+0.3$     &\phnn3.0\phn   &H,L    \\
20 21 44.201   &37 26 37.98    &\phnn2.4\phn   &\phn$+0.7$     &\phnn2.4\phn   &H,L    \\
\hline
\noalign{\smallskip}
\multicolumn{5}{l}{\phn \chtoh\ maser components}   \\
\hline
\noalign{\smallskip}
20 21 44.397   &37 26 48.14    &\phnn0.06      &\phn$+1.3$     &\phnn0.04      &1      \\
20 21 44.403   &37 26 48.15    &\phnn0.07      &\phn$-0.3$     &\phnn0.06      &1      \\
20 21 44.412   &37 26 48.04    &\phnn0.4\phn   &\phn$+1.3$     &\phnn0.4\phn   &1      \\
20 21 44.424   &37 26 47.96    &\phnn1.6\phn   &\phn$+0.5$     &\phnn2.4\phn   &1      \\
20 21 44.701   &37 26 41.30    &\phnn6.8\phn   &\phn$+3.8$     &\phnn5.6\phn   &2      \\
20 21 44.710   &37 26 41.47    &\phnn1.2\phn   &\phn$+4.2$     &\phnn1.0\phn   &2      \\
20 21 44.763   &37 26 42.17    &\phnn5.8\phn   &\phn$+3.2$     &\phnn6.7\phn   &3      \\
20 21 44.943   &37 26 56.28    &\phnn1.8\phn   &\phn$+0.5$     &\phnn1.2\phn   &4      \\
\hline\hline
\end{tabular}
\begin{list}{}{}
\item[$^\mathrm{a}$] Labels H and L (for water masers) refer to the VLA observations done with a spectral resolution of 0.33~\kms\ and 1.3~\kms, respectively. Numbers 1--4 (for methanol masers) refer to the four groups of methanol masers detected (see Figure~\ref{fg75:region} panel a)
\end{list}
\label{tg75:masers}
\end{table}

\subsection{\water\ and \chtoh\ maser emission} \label{sg75:resmaser}

Hofner \& Churchwell (1996) reported a cluster of water masers located 2$\arcsec$ southwest from the \uchii\ region, at the same position as the CORE source (as reported by Carral \et\ 1997). Our \water\ maser observations at 22.235~GHz have a double aim: to cross-calibrate the radio continuum data at 1.3~cm, and to observe the masers with higher angular resolution ($\sim$$0\farcs1$ versus $\sim$$0\farcs4$ of previous observations). Two different spectral resolutions ($\sim$0.3 and $\sim$1.3~\kms) were used to observe the \water\ maser emission, allowing us to look for maser components in a velocity range of $(-40,+40)$~\kms. In Figure~\ref{fg75:masers} we show the different \water\ maser positions (colored circles) overlaid on the 1.3~cm continuum image. We detected a total of 35 spots with velocities ranging from $-36$ to $+14$~\kms. In Table~\ref{tg75:masers} we list the position, intensity, velocity, and integrated intensity of each \water\ maser component, indicating in the last column at which spectral resolution the component was detected. The water masers appear to form an arc at a distance of $\sim$2\arcsec\ from the head of the cometary \uchii\ region (cf.\ Figure~\ref{fg75:region}), with only a few of them directly associated with the CORE-S continuum source. We note that there are no instrumental offsets to be considered between the 1.3~cm continuum and the \water\ maser images, since the observations were simultaneous. This arc distribution of the maser spots was also reported by Ando \et\ (2011) with VERA observations at 10~mas scales. In Figure~\ref{fg75:masers} we use different colors as an indicator of the velocity of the maser component. No clear velocity gradients are found in the \water\ maser emission, although most of them appear slightly redshifted with respect to the cloud velocity, $v_\mathrm{LSR}\approx0$~\kms\ (Olmi \& Cesaroni 1999; Codella \et\ 2010). In Section~\ref{sg75:discore} we discuss the possible association between water masers and the continuum emission.

Regarding the class~I methanol masers at 44.069~GHz, we detect four general locations of emission (with a total of 8 different components; see Figure~\ref{fg75:region} panel a). In Table~\ref{tg75:masers} we list the position, intensity, velocity, and integrated intensity for all the methanol maser components. None of the 44~GHz \chtoh\ spots appear to be directly associated with any of the radio continuum sources, but rather are located at a distance of 10\arcsec--20\arcsec\ (0.2--0.4~pc for a distance of 3.8~kpc) to the north-east of the three main YSOs. As in other star-forming regions (\eg\ Kurtz \et\ 2004), class~I methanol masers rarely coincide with other signposts of star formation (\eg\ \hii\ regions, OH masers, class~II methanol masers). Theoretical models of methanol masers suggest that the class~I masers arise in an environment where collisional processes dominate (Cragg \et\ 1992; Pratap \et\ 2008). The G75 star-forming region contains up to four different molecular outflows (Shepherd \et\ 1997), suggesting that the 44~GHz \chtoh\ masers could be pumped by collisions resulting from the molecular outflows. In particular, some of the methanol masers we detect are close to or within the red lobe of a molecular outflow seen in the CO\,(2--1) and $^{13}$CO\,(2--1) lines (S\'anchez-Monge 2011; see also Figure~\ref{fg75:IRall}). Furthermore, the velocities of these maser components are red-shifted (see Table~\ref{tg75:masers}) which would be expected if they are excited by collisions in the red-shifted lobe of the molecular outflow.

\subsection{Radio recombination line emission} \label{sg75:resrrls}

We used the OVRO interferometer to observe the radio recombination lines at 3~mm (H40$\alpha$) and 1~mm (H30$\alpha$) towards G75, to determine the contribution of the ionized gas component at millimeter wavelengths. However, we did not clearly detect either of these lines. The spectra are dominated by continuum emission coming from  either dust or ionized gas (see Section~\ref{sg75:seds}). Our $1\sigma$ rms noise levels for the line (continuum-free emission) are 10 and 15~m\jpb\ for the H40$\alpha$ and H30$\alpha$, respectively. Although we did not detect the radio recombination lines, the upper limits can be used to constrain the emission from ionized gas in the millimeter range (see Section~\ref{sg75:discore} for more details).

\begin{figure}[t!]
\begin{center}
\begin{tabular}[c]{c}
       \epsfig{file=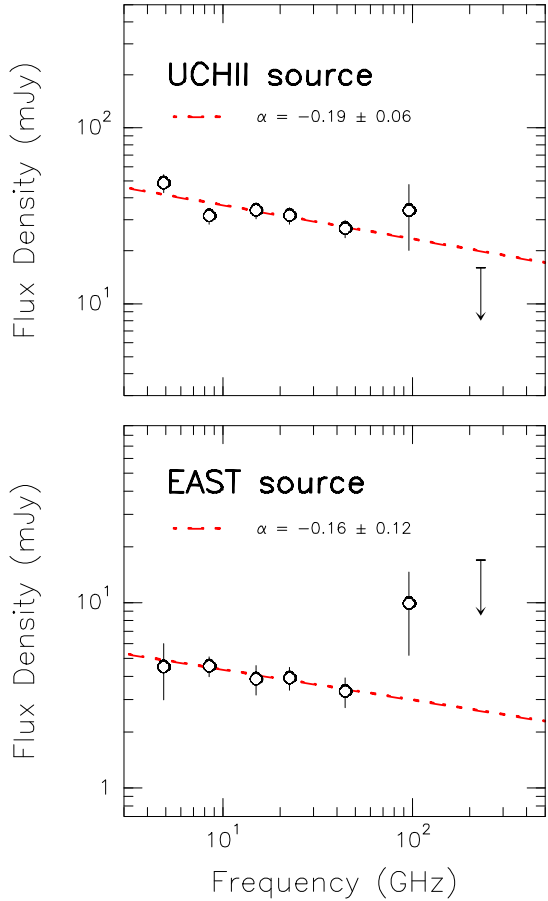, scale=0.8} \\
\end{tabular}
\caption{Spectral energy distributions for UCHII (top) and EAST (bottom) sources. Circles and upper limits correspond to the observational data from Table~\ref{tg75:rescont}. Red dashed lines: linear fit ($S_\nu\propto\nu^\alpha$) to the centimeter data (from 6 up to 0.7~cm; see Table~\ref{tg75:physpar}).}
\label{fg75:seds1}
\end{center}
\end{figure}

\begin{figure}[t!]
\begin{center}
\begin{tabular}[c]{c}
       \epsfig{file=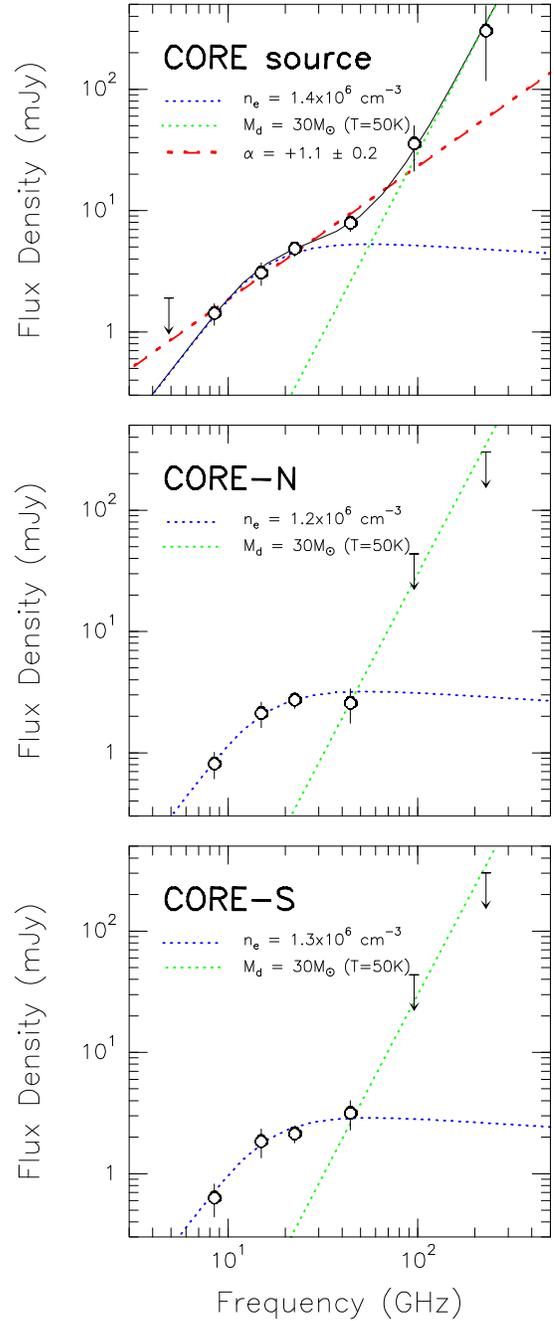, scale=0.8} \\
\end{tabular}
\caption{Flux density distributions for CORE (top), CORE-N (middle) and CORE-S (bottom) sources. Circles and upper limits correspond to the data from Table~\ref{tg75:rescont}. Red dashed lines: linear fit ($S_\nu\propto\nu^\alpha$) to the centimeter data (from 6 to 0.7~cm; see Table~\ref{tg75:physpar}). Blue dotted lines: homogeneous \hii\ region with an electron density specified in the panel. Green dotted lines: modified black body law for the dust envelope with a dust emissivity index of $\beta=1$, a source radius of $3$\arcsec, a dust temperature of 50~K, a dust mass of 30~\mo, and a dust mass opacity coefficient of 0.9~cm$^{2}$~g$^{-1}$ at 1.3~mm (Ossenkopf \& Henning 1994).}
\label{fg75:seds2}
\end{center}
\end{figure}

\section{Flux density distributions} \label{sg75:seds}

In Figures~\ref{fg75:seds1} and \ref{fg75:seds2} we show the flux density distributions (FDDs) of the radio continuum sources found towards G75. The UCHII source has a flat distribution, with a spectral index ($\alpha$; $S_\nu\propto\nu^\alpha$) of $-0.19\pm0.06$, typical of optically thin free-free emission. This FDD is well-fit by an optically thin \hii\ region with an electron density of $3.7\times10^{4}$~cm$^{-3}$, a diameter of $\sim$0.019~pc ($\sim$3800~AU; consistent with the observed deconvolved size), and ionized by a B0 type star. In Table~\ref{tg75:physpar} we list the main physical parameters of the cometary \uchii\ region. The emission detected at 3.1~mm and the upper limit at 1.3~mm are consistent with the millimeter continuum emission coming from ionized gas, suggesting that almost no dust emission is associated with the cometary \uchii\ region, in agreement with Riffel \& L\"udke (2010).

The eastern continuum source (EAST), barely resolved (cf.\ Figure~\ref{fg75:cont} and Table~\ref{tg75:rescont}), has a flat spectral index ($\alpha=-0.16\pm0.12$). Its FDD (see Figure~\ref{fg75:seds1}) can be fit, at centimeter wavelengths, by an optically thin \hii\ region with diameter $\sim$0.004~pc (900~AU; consistent with the deconvolved size listed in Table~\ref{tg75:rescont}) and an electron density of $1.3\times10^{5}$~cm$^{-3}$, requiring the equivalent of at least one B0.5 star to provide the ionizing photon flux. Its physical parameters are listed in Table~\ref{tg75:physpar}. At 3.1~mm there is an excess of continuum emission with respect to the optically thin \hii\ region assumption, probably coming from a dust envelope of $\sim$15--36~\mo\ (see last column in Table~\ref{tg75:physpar}). We refrain from fitting a dust envelope to the millimeter points of the FDD, because we have only an upper limit at 1.3~mm. Instead, we estimate an upper limit for the dust emissivity index $\beta<1.4$, from the 3~mm detection and the 1~mm upper limit.

\begin{table*}[ht!]
\caption{Physical parameters of the \hii\ regions and dust properties for the sources in G75.78+0.34}
\centering
\begin{tabular}{l c c c c c c c c c c c}
\noalign{\smallskip}
\hline\hline\noalign{\smallskip}
&\multicolumn{6}{c}{\hii\ region physical parameters\supa}
&
&\multicolumn{3}{c}{Dust\supb}
\\
\cline{2-7}\cline{9-11}\noalign{\smallskip}

&Diameter
&$EM$
&$n_\mathrm{e}$
&$M_\mathrm{i}$
&$\dot{N}_\mathrm{i}$
&Spectral
&
&$S_{\nu\mathrm{3mm}}^\mathrm{dust}$
&$S_{\nu\mathrm{1mm}}^\mathrm{dust}$
&$M_\mathrm{dust}$\supc
\\
Source
&(AU)
&(cm$^{-6}$~pc)
&(cm$^{-3}$)
&(\mo)
&(s$^{-1}$)
&Type
&
&(mJy)
&(mJy)
&(\mo)
\\
\noalign{\smallskip}
\hline\noalign{\smallskip}
\uchii\	&\phb\ 3800		&\phb\ $2.5\times10^{7}$	&\phb\ $3.7\times10^{4}$	&\phb\ $4.7\times10^{-3}$	&$4.2\times10^{46}$	&B0	&
		&7		&\ldots		&\ldots		\\
EAST		&\phb\ \phn900	&\phb\ $6.6\times10^{7}$	&\phb\ $1.3\times10^{5}$	&\phb\ $1.4\times10^{-4}$	&$4.4\times10^{45}$	&B0.5	&
		&8		&$<12$		&15--36 	\\
CORE		&\phn$<200$		&$>2.1\times10^{9}$		&$>1.4\times10^{6}$		&$<2.9\times10^{-5}$		&$1.1\times10^{46}$	&B0.5	&
		&30		&\phb\ 190	&30		\\
		&				&				&				&				&			&	&
		&5		&\phb\ 110	&17		\\
CORE-N	&\phn$<200$		&$>1.2\times10^{9}$		&$>1.2\times10^{6}$		&$<1.4\times10^{-5}$		&$4.2\times10^{45}$	&B0.5	&
		&$<40$		&$<200$		&$<30$		\\
CORE-S	&\phn$<200$		&$>1.4\times10^{9}$		&$>1.3\times10^{6}$		&$<1.3\times10^{-5}$		&$4.1\times10^{45}$	&B0.5	&
		&$<40$		&$<200$		&$<30$		\\
\hline\hline
\end{tabular}
\begin{list}{}{}
\item[\supa] Physical parameters of the \hii\ regions assuming homogeneous ionized gas. For the EAST and \uchii\ sources the parameters were obtained by fitting an optically thin \hii\ region at all centimeter wavelengths, and for the CORE, CORE-N and CORE-S sources the parameters correspond to the fit of an homogeneous \hii\ region at all centimeter wavelengths (see Figure~\ref{fg75:seds1}). The fit of an homonegeous \hii\ region is obtained from the expression
\begin{equation}
S_\nu = B_\nu(T_\mathrm{e})~(1-e^{-\tau_{ff}(\nu)})~\Omega_\mathrm{source} = \frac{2h\nu^3}{c^2}~\frac{1}{e^{h\nu/kT_\mathrm{e}}-1}~(1-e^{-\tau_{ff}(\nu)})~\Omega_\mathrm{source},
\end{equation}
where $S_\nu$ is the flux density at frequency $\nu$, $B_\nu(T_\mathrm{e})$ is the Planck function corresponding to the electronic temperature $T_\mathrm{e}$ assumed to be $10^4$~K, $\Omega_\mathrm{source}$ is the source solid angle, and $\tau_{ff}(\nu)$ is the optical depth defined by $\tau_{ff}(\nu)\approx0.08235\big[\frac{EM}{\mathrm{cm}^{-6}~\mathrm{pc}}\big]~\big[\frac{T_\mathrm{e}}{\mathrm{K}}\big]^{-1.35}~\big[\frac{\nu}{\mathrm{GHz}}\big]^{-2.1}$ (Altenhoff \et\ 1960). The free parameters in the fit are the emission measure and the size of the source. The spectral type is determined from Panagia (1973) using the number of ionized photons, $\dot{N}_\mathrm{i}$.
\item[\supb] $S_{\nu\mathrm{3mm}}^\mathrm{dust}$ and $S_{\nu\mathrm{1mm}}^\mathrm{dust}$ correspond to the (dust) millimeter continuum flux density after subtracting the contribution from free-free emission (from the fits shown in Figure~\ref{fg75:seds1}). For the CORE source, the first value assumes the centimeter emission is described by an homogeneous \hii\ region, and the second value assumes the centimeter emission is fitted with a power law fit ($S_\nu\propto\nu^\alpha$) with $\alpha=1.1$.
\item[\supc] Dust and gas mass estimated from the millimeter emission (after subtracting the contribution of the ionized gas). For the \uchii\ source, all the millimeter continuum emission is thermal ionized gas emission. For the EAST source we assumed a dust emissivity index of 1.5, a dust mass opacity coefficient of 0.9~cm$^2$~g$^{-1}$ at 1.3~mm (Ossenkopf \& Henning 1994), and a dust temperature of 50 and 20~K, respectively. For CORE, CORE-N and CORE-S sources we provide the mass estimated assuming a dust emissivity index of 1, a dust mass opacity coefficient of 0.9~cm$^2$~g$^{-1}$ at 1.3~mm, and a dust temperature of 50~K. The upper limits are due to the low angular resolution at millimeter wavelengths that does not allow to resolve the CORE-N and CORE-S sources.
\end{list}
\label{tg75:physpar}
\end{table*}

Finally, we construct the FDD for the CORE source, first including all the emission and then for CORE-N and CORE-S separately (Figure~\ref{fg75:seds2}). At centimeter wavelengths, the emission is partially optically thick, with a spectral index of $+1.1\pm0.2$ for the total emission of the CORE. The major difference between the FDDs of CORE, CORE-N and CORE-S  is the flux density at 7~mm. However, in all cases the centimeter emission can be well-fit by an \hii\ region with an electron density of $\sim$10$^6$~cm$^{-3}$ and a size $<0.001$~pc ($<200$~AU; see Table~\ref{tg75:physpar}). An \hii\ region with these properties has a spectral turnover in the centimeter regime. At millimeter wavelengths, where we spatially resolve the contributions of CORE-N and CORE-S, the emission is dominated by hot dust; we estimate an envelope mass of 30~\mo\ and a temperature of 50~K (see Table~\ref{tg75:physpar}).

\begin{figure*}[htp!]
\begin{center}
\begin{tabular}[b]{c}
	\epsfig{file=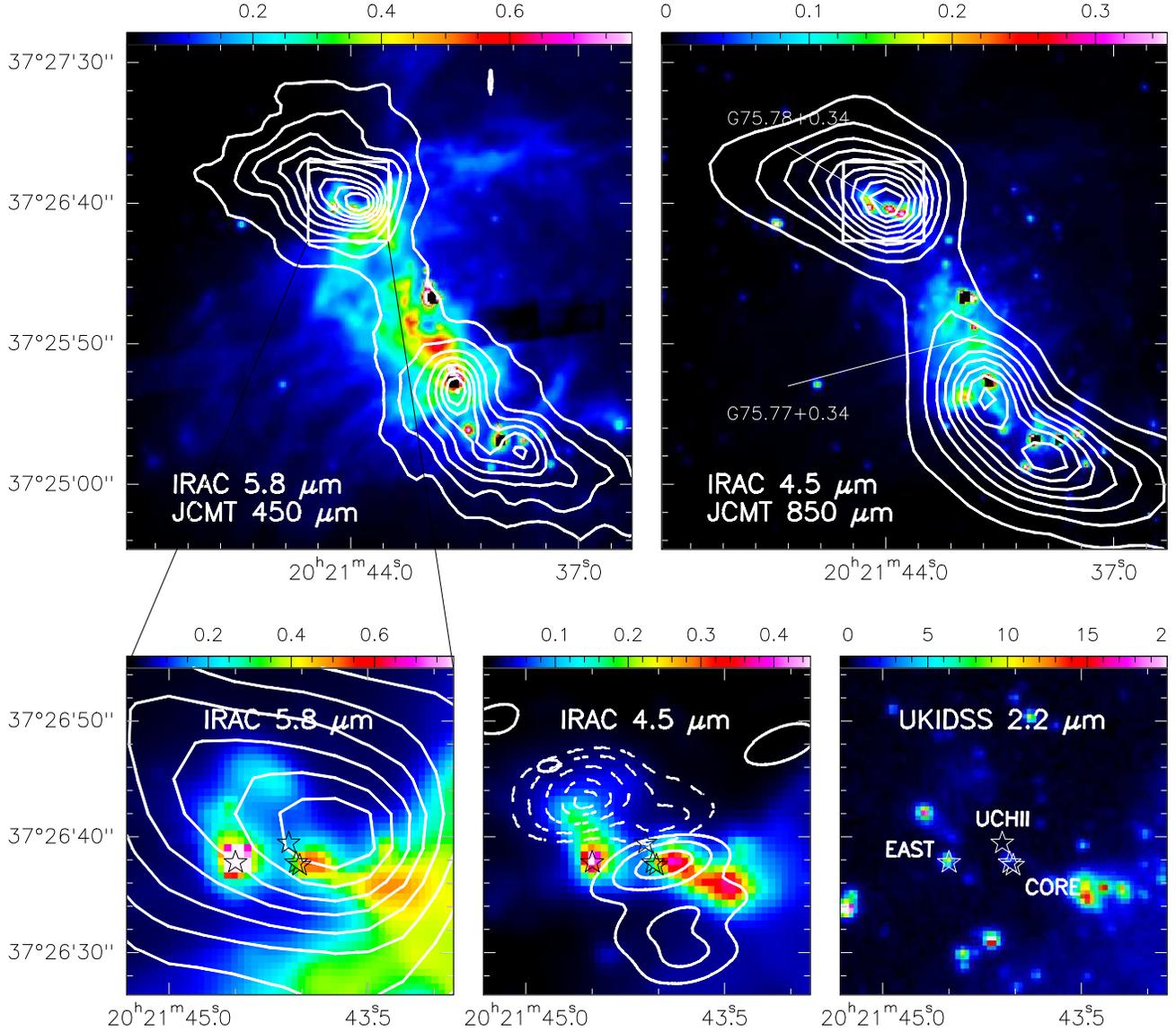, scale=0.88, angle=0} \\
\end{tabular}
\caption{Submillimeter (JCMT) and infrared (\emph{Spitzer} and UKIDSS) images of G75. {\bf Top-left}: 450~$\mu$m JCMT continuum image (white contours, levels start at 10\% increasing in steps of 10\% of the peak intensity 99.6~Jy) overlaid on the 5.8~$\mu$m IRAC-Spitzer image. {\bf Top-right}: 850~$\mu$m JCMT continuum image (white contours, levels start at 10\% increasing in steps of 10\% of the peak intensity 10.3~Jy), overlaid on the 4.5~$\mu$m IRAC-Spitzer image. The white box shows the region zoomed in the bottom panels. The large infrared and submillimeter source toward the southwest of the images corresponds to the \hii\ region G75.77+0.34. A dark region is seen toward the northeast in the infrared images, coincident with the strong submillimeter emission, and associated with the G75 star forming complex. The strongest sources in the region, including the large \hii\ region G75.77+0.33, appear saturated (black pixels) in the Spitzer images. {\bf Bottom}: Close-up view of the region studied with interferometers showing the 5.8~$\mu$m, 4.5~$\mu$m, and 2.2~$\mu$m images. The white contours in the lower left panel show the 450~$\mu$m JCMT continuum emission as in the top panel. The solid/dashed white contours in the bottom middle panel show the blue/red-shifted CO\,(2--1) emission (S\'anchez-Monge 2011; the contours are 10, 30, 70, 120 and 150~\jpb~\kms) tracing the molecular outflow with a direction similar to the elongation seen at 4.5~$\mu$m. The white/black stars mark the positions of the radio continuum sources: UCHII, EAST and CORE-N and CORE-S (see Table~\ref{tg75:rescont}). No infrared emission is associated with the UCHII source, while strong emission is coincident with the EAST source (see main text for discussion). In all panels, the units of the color scales are $10^3$~MJy~sr$^{-1}$.}
\label{fg75:IRall}
\end{center}
\end{figure*}

\section{Nature of the ionized gas emission} \label{sg75:disc}

We report the detection of four distinct radio continuum sources toward the high-mass star forming region G75: UCHII, EAST, CORE-N and CORE-S (the latter two correspond to the CORE source for angular resolutions coarse than $0\farcs3$). In this section we discuss the nature of these four sources.

\subsection{UCHII and EAST: two ultracompact \hii\ regions} \label{sg75:disuchii}

The UCHII and EAST sources can be well-characterized as  homogeneous \hii\ regions with sizes (3800~AU and 900~AU) and electron densities ($3.7\times10^4$~cm$^{-3}$ and $1.3\times10^5$~cm$^{-3}$) characteristic of small ultracompact \hii\ regions (\eg\ Kurtz 2005).  They are probably ionized by B0 and B0.5 ZAMS stars, respectively, and are optically thin at centimeter wavelengths (\ie\ spectral index of $-0.1$; see Table~\ref{tg75:physpar}). Despite these similarities, the two sources are distinctly different in their infrared emission.  In particular, there is substantial IR dust emission associated with the EAST source but not with the UCHII source. In Figure~\ref{fg75:IRall} we show infrared and submillimeter images of the G75 star-forming complex. Infrared emission between 2.2 and 8.0~$\mu$m is clearly associated with the EAST source, while no infrared emission at these wavelengths is associated with the cometary UCHII source (cf.\ close-up views in Figure~\ref{fg75:IRall}).  The absence of infrared emission from the UCHII source is unusual; ultracompact \hii\ regions typically present substantial IR emission (Hoare et al. 2007).  A possible explanation is that the \uchii\ region may be partially obscured by a cloud of gas and dust that absorbs the near- and mid-infrared emission.  In fact, such a cloud is seen in emission at submillimeter wavelengths (cf.\ emission at 450 and 850~$\mu$m in Figure~\ref{fg75:IRall}; Di Francesco \et\ 2008), and in absorption at infrared wavelengths (cf.\ obscured region in the infrared maps in Figure~\ref{fg75:IRall}), and its peak falls very close to the UCHII source. In addition, if the EAST source were located closer to the border of the dust cloud (seen in the 450~$\mu$m and 850~$\mu$m images), its IR emission would be less obscured than that of the UCHII source, resulting in the different infrared properties of the two sources. The non-detection of UCHII at 1.3~mm (see Fig.~\ref{fg75:contmm}) does not contradict this scenario; as we noted in Sect.~\ref{sg75:resmm}, the non-dection is probably due to inadequate sensitivity.

\subsection{CORE: hypercompact \hii\ regions or shocks?} \label{sg75:discore}

Regarding the emission from the CORE (N and S) source, we examine several models to explain the origin of the ionized gas emission. Franco \et\ (2000) modeled the CORE source as an \hii\ region with a density distribution $n_\mathrm{e}\propto r^{-4}$, motivated by optically thick centimeter continuum emission at frequencies up to $\sim$100~GHz. Our new, higher resolution observations suggest that the emission becomes optically thin at frequencies $\sim$30~GHz. Figure~\ref{fg75:seds2} suggests that a  homogeneous density distribution can account for the emission from the N and S sources. Furthermore, the RRL upper limits are also consistent with optically thin emission. Assuming optically thin emission for line and continuum, and $3\sigma$ upper limits for H30$\alpha$ and H40$\alpha$ (\ie\ 30 and 45~m\jpb, respectively), we can calculate the expected continuum emission from ionized gas with (Rohlfs \& Wilson 2004)
\begin{equation}
\bigg[\frac{S_\mathrm{L}}{S_\mathrm{C}}\bigg] = 
6940 \bigg[\frac{\Delta v}{\mathrm{km~s}^{-1}}\bigg]^{-1}
\bigg[\frac{T_\mathrm{e}}{\mathrm{K}}\bigg]^{-1.15}
\bigg[\frac{\nu}{\mathrm{GHz}}\bigg]^{1.1},
\end{equation}
where $S_\mathrm{L}$ is the line flux density, $S_\mathrm{C}$ is the continuum flux density, $\Delta v$ is the width of the radio recombination line (assumed to be 30~\kms; Kurtz 2005), $T_\mathrm{e}$ is the electron temperature assumed to be $10^4$~K, and $\nu$ is the frequency of the line. We find that the continuum flux density of ionized gas at 3 and 1~mm should be $<$30 and $<$18~mJy, respectively. These values are in agreement with the homogeneous \hii\ region fits given in Section~\ref{sg75:seds}, for which the expected flux densities at 3 and 1~mm are $\sim$3~mJy (\ie\ $\ll$18~mJy). Thus, we conclude that the ionized gas emission becomes optically thin for frequencies $>$30~GHz.  This new result, compared to Franco et al. (2000), arises because we resolve the CORE source into two components.

From the FDD analysis (see Section~\ref{sg75:seds}), both sources (CORE-N and S) appear to be hypercompact \hii\ regions ionized by B0.5 ZAMS stars, and separated by a distance of $0\farcs36$ (1400~AU at a distance of 3.8~kpc). This scenario suggests a wide, massive binary system (Sana \& Evans 2011). We searched the literature for similar systems, with two close massive radio continuum sources, and found several examples: in G31.41+0.31 there is a system of two centimeter continuum sources separated by $0\farcs19$ (1500~AU at 7.9~kpc; Araya \et\ 2008); in G10.47+0.03 there are two similar sources separated by $0\farcs53$ (5700~AU at 10.8~kpc; Cesaroni \et\ 2010); and in DR21(OH) a similar system is separated by $0\farcs45$ (900~AU at 2~kpc; Araya \et\ 2009). Thus, it seems that it is not unusual to detect double radio continuum sources, separated by $\sim$2000~AU, in massive star-forming regions. 

It should be noted that the radio continuum emission from some of the previously listed binary systems is not always interpreted as arising from photo-ionized \hii\ regions. Araya \et\ (2009) propose that the radio continuum emission in DR(21)OH is free-free radiation originating in interstellar shocks, following the theoretical development of Ghavamian \& Hartigan (1998). In this scenario, the free-free emission arises from shock-ionized, rather than photo-ionized, gas, being comparable to the Herbig-Haro objects seen in the optical.

We note that the infrared emission at 4.5~$\mu$m (see Figure~\ref{fg75:IRall}) shows an elongated structure `emanating' from the position of the CORE source. The fluxes of this structure in the four IRAC/\emph{Spitzer} bands are 4~mJy at 3.6~$\mu$m, 52~mJy at 4.5~$\mu$m, 96~mJy at 5.8~$\mu$m, and 115~mJy at 8.0~$\mu$m, and thus it fulfills the criteria\footnote{Chambers \et\ (2009) classified an extended mid-infrared source as a `green fuzzy' according to the following color criteria: (1) the 4.5~$\mu$m to 3.6~$\mu$m flux ratio is $\ge$1.8, (2) the 4.5~$\mu$m to 5.8~$\mu$m flux ratio is $\ge$0.40, and (3) the 4.5~$\mu$m to 8.0~$\mu$m flux ratio is $\ge$0.45.} defined by Chambers \et\ (2009) to be classified as a `green fuzzy' (see also Cyganowski \et\ 2008). The excess 4.5~$\mu$m emission can be produced by shocks associated with outflows, scattered continuum in outflow cavities, obscuration affecting the emission at 3.6~$\mu$m, or fluorescence H$_2$ emission (\eg\ Noriega-Crespo \et\ 2004; Smith \et\ 2006; Qiu \et\ 2008; De Buizer \& Vacca 2010; Takami \et\ 2010; Varricatt 2011; Simpson \et\ 2012; Lee \et\ 2012; Takami \et\ 2012), and only spectroscopic studies in the mid-infrared can clearly determine the origin of the 4.5~$\mu$m excess. However, in G75, the northeast-southwest direction of the 4.5~$\mu$m elongated structure is consistent with the direction of the molecular outflow likely associated with the CORE source (see the solid/dashed contours in the bottom middle panel of Figure~\ref{fg75:IRall}, S\'anchez-Monge 2011) which is also consistent with the direction of the outflow reported by Riffel \& L\"udke (2010) at larger scales, thus favoring the interpretation of the 4.5~$\mu$m emission arising from shocks. In addition, the spatial distribution of water masers (Figure~\ref{fg75:masers}) is also indicative of shocks and jets, as found in other star forming regions (\eg\ IRAS~20126+4104: Hofner \et\ 2007). A milliarcsecond kinematical study of water masers in G75.78+0.34 (see Figure~3 in Ando \et\ 2011) is consistent with a shock interpretation for the centimeter continuum emission. The group of masers associated with CORE-S have velocities mainly in the southwest direction (similar to the elongated structure seen at 4.5~$\mu$m), while the two groups of maser spots located to the northwest and southeast of CORE-S show a velocity field suggesting expansion. Together with the arc morphology of all the maser spots this is suggestive of a bow-shock, implying that at least the CORE-S emission might arise from shocks. 


A stellar jet/outflow with a mass-loss rate larger than $4~10^{-6}$~\mo~yr$^{-1}$ and a terminal velocity of order several hundred km~s$^{-1}$, shocking the ambient medium, would have sufficient energy to ionize, heat, and move the gas of both the CORE-N and CORE-S sources (see Appendix~\ref{appen}). Although the cooling time (and hence recombination) of the high density gas is very short (Franco \et\ 1994), the jet/outflow would continuously re-ionize the material. In this case, the exciting source would be located between the two radio continuum sources, with its own \hii\ region quenched by the high density medium. I.e., a jet/outflow with such a high mass-loss rate would be neutral because the ionization front would be trapped very close to the star.  The possibility of a neutral jet/outflow shocking the ambient gas cannot presently be ruled out because our observations are sensitive to ionized gas.  This scenario can be tested with future centimeter continuum observations with the JVLA, to determine if CORE-S (and the water maser spots) have been displaced with respect to the other continuum sources. Assuming a jet velocity of order several hundred \kms, relative motions could be detected within the next five years.

As noted in Section~\ref{sg75:intro}, other mechanisms might also cause the centimeter emission from CORE-N and S. However,we consider most of these to be unlikely candidates. Equatorial winds (Hoare 2006) or thermal radio jets (Anglada 1996) typically predict constant spectral indices of $+0.6$ and a relation between the source size and the frequency. Our observations indicate spectral indices $\sim$+1; the uncertainties in the deconvolved source sizes are too large to determine a size--frequency dependence.  A photoevaporated disk model has been proposed to explain the continuum emission of double-peaked sources with an hour-glass morphology (Lugo \et\ 2004). Although the CORE N and S peaks might be fit as a photoevaporated disk, recent studies indicate that the peaks from photoevaporated disks should change their separation with frequency, moving closer at higher frequencies (Tafoya \et\ 2004). The peak positions of N and S  do not change between 3.6 and 0.7~cm to a precision of $\sim$$0\farcs02$. 

Summarizing, the centimeter continuum emission from CORE N and S can be well fitted by homogeneous \hii\ regions, each one photo-ionized by a B0.5 ZAMS star. Alternatively, free-free radiation from shock-ionized gas resulting from the interaction of a jet/outflow system with the surrounding environment appears to be a viable scenario as well. Other possible mechanisms are deemed unlikely.  

\section{Conclusions} \label{scon}

We have carried out subarcsecond angular resolution observations with the VLA in the centimeter continuum and in \water\ and \chtoh\ maser emission toward the massive star forming complex G75.78+0.34. We have complemented these observations with OVRO and SMA millimeter continuum and radio recombination line observations, and submillimeter/infrared data from on-line databases. Our conclusions can be summarized as follows:

\begin{list}{}{}

\item[1.] Our radio continuum data reveal centimeter continuum emission at all wavelengths, with the emission coming from four distinct compact sources: UCHII, EAST, CORE-N and CORE-S. The two last sources appear unresolved (CORE source) when observed with angular resolutions $\gtrsim$$0\farcs3$.

\item[2.] The strongest source, UCHII, is a cometary \uchii\ region with a size of 0.02~pc (3800~AU), an electron density of $3.7\times10^{4}$~cm$^{-3}$ and is ionized by the equivalent of a B0 ZAMS star. The EAST source, located $\sim\!6$\arcsec\ to the east of UCHII, is also an \uchii\ region with an electron density of $1.3\times10^{5}$~cm$^{-3}$, ionized by a B0.5 equivalent ZAMS star, but with a smaller (barely resolved) size of 0.004~pc (900~AU). The millimeter continuum emission associated with these two \uchii\ regions has different origins: for the UCHII source it probably traces the ionized gas, while for the EAST source there is a millimeter excess, suggesting that this source is still embedded in a compact dust clump also detectable at mid-infrared wavelengths, with a mass of 15--36~\mo.  We suggest that the non-detection of thermal dust emission in the UCHII source results from foreground extinction, not from an absence of warm dust.

\item[3.] The CORE-N and S sources are located 2\arcsec\ to the southwest of the cometary arc of the UCHII source, and are associated with water maser and dense molecular gas emission. The two sources are very close to one another ($0\farcs36$, 1400~AU) and have similar properties: flux density increasing with frequency, and unresolved emission. The free-free continuum emission may be produced by two hypercompact \hii\ regions ionized by B0.5 ZAMS stars or by the collision between a jet/outflow and the surrounding environment. 
A dust clump of $\sim\!30$ ~\mo\ associated with the CORE source is detected at millimeter wavelengths.

\item[4.] We have also reported high angular resolution observations of \water\ and class~I \chtoh\ maser emission. The class~I methanol maser spots are found far from any of the radio continuum sources, and are probably excited by collisional processes due to the presence of multiple molecular outflows. The water maser spots appear close to the CORE-N and S sources, with only a few of them spatially coincident with the S source. The spatial distribution of the water masers and the kinematics observed at milliarcsecond angular resolution, suggest that they could originate in a shock between a jet/outflow and the local surroundings. In addition, extended 4.5~$\mu$m emission likely tracing shocked gas could be associated with a jet driven by the CORE source.

\end{list}

In summary, we have characterized the radio-continuum emission toward the massive star forming complex G75.78+0.34, revealing the presence of four sources: two ultracompact \hii\ regions (named UCHII and EAST) ionized by B0--B0.5 ZAMS stars, one with an extended and cometary shape and the other being barely resolved; and two compact sources (named CORE-N and CORE-S) associated with at least one massive protostar embedded in a dense and massive envelope and with hints of driving an outflow.
\vspace{0.2cm}

\acknowledgments
\begin{small}
The authors are grateful to the anonymous referee for valuable comments. \'A.S.-M., A.P.\ and R.E.\ are partially supported by the Spanish MICINN grants AYA~2008-06189-C03 and AYA~2011-30228-C03 (co-funded with FEDER funds). \'A.S.-M.\ and S.L.\ acknowledges support by PAPIIT-UNAM IN100412 and CONACyT 146521. S.K.\ acknowledges support from DGAPA, UNAM, project IN101310. A.P.\ is supported by a JAE-Doc CSIC fellowship co-funded with the European Social Fund under the program `Junta para la Ampliaci\'on de Estudios', by the Spanish MICINN grant AYA2011-30228-C03-02 (co-funded with FEDER funds) and by the AGAUR grant 2009SGR1172 (Catalonia). G.G.\ acknowledges support from CONICYT project PFB-06. This publication makes use of data products from the Two Micron All Sky Survey, which is a joint project of the University of Massachusetts and the Infrared Processing and Analysis Center/California Institute of Technology, funded by the National Aeronautics and Space Administration and the National Science Foundation.
\end{small}

\appendix
\section{Energy budget} \label{appen}

Consider a star that ejects into a cloud a bipolar jet/outflow with a mass-loss rate $\dot{M}_\mathrm{j}$ and a jet velocity $v_\mathrm{j}$. The jet  will collide with the ambient material in two opposite directions and will produce two shocked ionized regions. The luminosity of the jet, $L_\mathrm{j}$, is given by
\begin{equation}\label{eq:Lj}
L_\mathrm{j} = \frac{1}{2}\,\dot{M}_\mathrm{j}\,v_\mathrm{j}^2 = 
7.9\times10^{35}\,\bigg[\frac{\dot{M}_\mathrm{j}}{10^{-5}~\mathrm{M}_\sun~\mathrm{yr}^{-1}}\bigg]\,\bigg[\frac{v_\mathrm{j}}{500~\mathrm{km~s}^{-1}}\bigg]^2~\mathrm{erg~s}^{-1}.
\end{equation}
The shocked ionized gas is  post shock ambient gas pushed and heated by the stellar bipolar jet. For simplicity, we will assume both shocked ionized regions have the same physical characteristics. Then, the minimum luminosity required to keep the post shock gas ionized is
\begin{equation}\label{eq:Li}
L_\mathrm{i} = E_0\,\dot{N}_\mathrm{i} = 
2.2\times10^{35}\,\bigg[\frac{\dot{N}_\mathrm{i}}{10^{46}~\mathrm{s}^{-1}}\bigg]~\mathrm{erg~s}^{-1},
\end{equation}
where $E_0=13.6$~eV is the ionization energy of the hydrogen atom and $\dot{N}_\mathrm{i}$ is the rate of ionizing photons inferred from the radio continuum emission of the ionized region. Furthermore, the energy lost by radiation from each region is 
\begin{equation}\label{eq:Lrad}
L_\mathrm{rad}\,\sim\,\Lambda(T)\,V_\mathrm{i} = 
4.2\times10^{34}\,\bigg[\frac{n_\mathrm{e}}{10^6~\mathrm{cm}^{-3}}\bigg]^2\,\bigg[{\frac{r_\mathrm{i}}{100~\mathrm{AU}}}\bigg]^3~\mathrm{erg~s}^{-1},
\end{equation}
where we assume a cooling rate per unit volume $\Lambda(10^4~\mathrm{K})/(n_\mathrm{e}~n_\mathrm{p})\sim 3\times10^{-24}$~erg~cm$^{3}$~s$^{-1}$ (Osterbrock 1989), a volume of each region with radius $r_\mathrm{i}$, $V_\mathrm{i}=4\pi r_\mathrm{i}^3/3$, and assume that the electron density is equal to the proton density, $n_\mathrm{e}=n_\mathrm{p}$. Also, we estimate the kinetic luminosity of each ionized region averaged over its lifetime as
\begin{equation}\label{eq:Lk}
L_\mathrm{K}\,\sim\,\frac{M_\mathrm{i}\,v_\mathrm{i}^2}{2\,\tau_\mathrm{cross}} = 
1.5\times10^{34}\,\bigg[\frac{n_\mathrm{e}}{10^6~\mathrm{cm}^{-3}}\bigg]\,\bigg[\frac{r_\mathrm{i}}{100~\mathrm{AU}}\bigg]^3\,\bigg[\frac{v_\mathrm{i}}{200~\mathrm{km~s}^{-1}}\bigg]^2\,\bigg[\frac{\tau_\mathrm{cross}}{10~\mathrm{yr}}\bigg]^{-1}~\mathrm{erg~s}^{-1},
\end{equation}
where the mass of ionized gas is $M_\mathrm{i}=m_\mathrm{H}~n_\mathrm{p}~V_\mathrm{i}$, the velocity is $v_\mathrm{i}$, and $\tau_\mathrm{cross}$ is the crossing time that can be estimated from $\tau_\mathrm{cross}=l/v_\mathrm{i}$, where $l$ is the distance from the ionized region to the star emitting the jet.

We consider now the particular case of the CORE-N and CORE-S sources. From Table~\ref{tg75:physpar} we take $n_\mathrm{i}\sim 10^6$~cm$^{-3}$, $r_\mathrm{i}\sim 100$~AU, and $\dot{N}_\mathrm{i}\sim 4\times10^{45}$~s$^{-1}$. We assume that each core is the ionized working surface (WS) of the jet moving into the cloud. For strong shocks, the speed of each WS is given by ram pressure balance as $v_\mathrm{i} = \beta v_\mathrm{j}/(1+\beta)$, where $\beta=\sqrt{\rho_\mathrm{j}/\rho_\mathrm{a}}$ is the square root of the ratio of the jet and ambient densities (\eg\ Raga \et\ 1990). For large jet mass-loss rates, close to the emitting source, we assume $\beta\sim 1$. Then, for a jet velocity $v_\mathrm{j}\sim 500$~\kms\ (of the order of the escape speed of a 10~\mo\ star assumed to be the source of the jet) the shocked ionized gas moves with a speed, $v_\mathrm{i}\sim v_\mathrm{j}/2\sim 250$~\kms. Because the two continuum sources are separated by $\sim 1700$~AU, we assume a distance to the jet source in between CORE-N and CORE-S, $l\sim 850$~AU. Then, the resulting crossing time for both ionized regions is $\tau_\mathrm{cross}\sim 16.2$~yr. Introducing these values in Eq.~\ref{eq:Li}, \ref{eq:Lrad}, and \ref{eq:Lk}, we obtain for each shocked ionized region $L_\mathrm{i}\approx8.8\times10^{34}$~erg~s$^{-1}$, $L_\mathrm{rad}\approx4.2\times10^{34}$~erg~s$^{-1}$, and $L_\mathrm{K}\approx1.5\times10^{34}$~erg~s$^{-1}$.


Finally, one can now estimate the minimum mass-loss rate and velocity of a jet necessary to ionized, heat, and move the CORE-N and CORE-S sources. The energy budget has to fulfill the relation $L_\mathrm{j} > 2(L_\mathrm{i}+L_\mathrm{rad}+L_\mathrm{K})$. From Eq.~\ref{eq:Lj}, the minimum mass-loss rate of the bipolar jet/outflow that heats and moves the ionized the regions would be $\dot{M}_\mathrm{j} > 3.7\times10^{-6}$~\mo~yr$^{-1}$. 




\begin{references} 

\reference{} Altenhoff, W., Mezger, P.~G., Wendker, H., \& Westerhout, G.\ 1960, Veroff.\ Univ.\ Sternwarte Bonn, 59, 48
\reference{} Ando, K., Nagayama, T., Omodaka, T., et al.\ 2011, \pasj, 63, 45 
\reference{} Anglada, G.\ 1996, Radio Emission from the Stars and the Sun, 93, 3 
\reference{} Araya, E.~D., Kurtz, S., Hofner, P., \& Linz, H.\ 2009, \apj, 698, 1321 
\reference{} Araya, E., Hofner, P., Kurtz, S., Olmi, L., \& Linz, H.\ 2008, \apj, 675, 420 
\reference{} \'Avalos, M., \& Lizano, S.\ 2012, \apj, 751, 63
\reference{} Beuther, H., \& Shepherd, D.\ 2005, Cores to Clusters: Star Formation with Next Generation Telescopes, 105 
\reference{} Bisschop, S.~E., Fuchs, G.~W., Boogert, A.~C.~A., van Dishoeck, E.~F., \& Linnartz, H.\ 2007, \aap, 470, 749 
\reference{} Briggs, D.\ 1995, PhD Thesis, New Mexico Inst. of Mining and Technology
\reference{} Campbell, M.~F., Niles, D., Nawfel, R., et al.\ 1982, \apj, 261, 550
\reference{} Carral, P., Kurtz, S.~E., Rodriguez, L.~F., de Pree, C., \& Hofner, P.\ 1997, \apjl, 486, L103 
\reference{} Cesaroni, R., Hofner, P., Araya, E., \& Kurtz, S.\ 2010, \aap, 509, A50 
\reference{} Chambers, E.~T., Jackson, J.~M., Rathborne, J.~M., \& Simon, R.\ 2009, \apjs, 181, 360 
\reference{} Codella, C., Cesaroni, R., L{\'o}pez-Sepulcre, A., Beltr{\'a}n, M.~T., Furuya, R., \& Testi, L.\ 2010, \aap, 510, A86 
\reference{} Cragg, D.~M., Johns, K.~P., Godfrey, P.~D., \& Brown, R.~D.\ 1992, \mnras, 259, 203 
\reference{} Cyganowski, C.~J., Whitney, B.~A., Holden, E., et al.\ 2008, \aj, 136, 2391 
\reference{} De Buizer, J.~M., \& Vacca, W.~D.\ 2010, \aj, 140, 196 
\reference{} Dent, W.~R.~F., MacDonald, G.~H., \& Andersson, M.\ 1988, \mnras, 235, 1397
\reference{} Di Francesco, J., Johnstone, D., Kirk, H., MacKenzie, T., \& Ledwosinska, E.\ 2008, \apjs, 175, 277 
\reference{} Eisloffel, J., Mundt, R., Ray, T.~P., \& Rodriguez, L.~F.\ 2000, Protostars and Planets IV, ed. V. Mannings, A. P. Boss, \& S. S. Russell (Tucson, AZ: Univ. Arizona Press), 815
\reference{} Elld{\'e}r, J., R{\"o}nn{\"a}ng, B., \& Winnberg, A.\ 1969, \nat, 222, 67
\reference{} Feigelson, E.~D., \& Montmerle, T.\ 1999, \araa, 37, 363 
\reference{} Franco, J., Kurtz, S., Hofner, P., Testi, L., Garc{\'{\i}}a-Segura, G., \& Martos, M.\ 2000, \apjl, 542, L143
\reference{} Franco, J., Miller, W. W., III, Arthur, S. J., Tenorio-Tagle, G., \& Terlevich, R.\ 1994, \apj, 435, 805
\reference{} Garay, G.\ 1987, RMxAA, 14, 489
\reference{} Ghavamian, P., \& Hartigan, P.\ 1998, \apj, 501, 687 
\reference{} Hanson, M.~M., Luhman, K.~L., \& Rieke, G.~H.\ 2002, \apjs, 138, 35
\reference{} Hoare, M.~G.\ 2006, \apj, 649, 856 
\reference{} Hoare, M.~G., Kurtz, S.~E., Lizano, S., Keto, E., \& Hofner, P.\ 2007, Protostars and Planets V, ed. B. Reipurth, D. Jewitt, \& K. Keil (Tucson, AZ: Univ. Arizona Press), 181
\reference{} Hofner, P., \& Churchwell, E.\ 1996, \aaps, 120, 283 
\reference{} Hollenbach, D., Johnstone, D., Lizano, S., \& Shu, F.\ 1994, \apj, 428, 654
\reference{} Keto, E.\ 2007, \apj, 666, 976 
\reference{} Keto, E.\ 2003, \apj, 599, 1196
\reference{} Keto, E.\ 2002, \apj, 568, 754
\reference{} Klaassen, P.~D., \& Wilson, C.~D.\ 2007, \apj, 663, 1092 
\reference{} Kurtz, S.\ 2005, Astrochemistry: Recent Successes and Current Challenges, 231, 47
\reference{} Kurtz, S., Hofner, P., \& {\'A}lvarez, C.~V.\ 2004, \apjs, 155, 149 
\reference{} Lee, H.-T., Takami, M., Duan, H.-Y., et al.\ 2012, \apjs, 200, 2
\reference{} Lugo, J., Lizano, S., \& Garay, G.\ 2004, \apj, 614, 807
\reference{} Marti, J., Rodriguez, L.~F., \& Reipurth, B.\ 1993, \apj, 416, 208
\reference{} Matthews, H.~E., Goss, W.~M., Winnberg, A., \& Habing, H.~J.\ 1973, \aap, 29, 309
\reference{} Matthews, H.~E., Goss, W.~M., Winnberg, A., \& Habing, H.~J.\ 1977, \aap, 61, 261
\reference{} Matthews, N., Anderson, M., \& MacDonald, G.~H.\ 1986, \aap, 155, 99
\reference{} Nagayama, T., \& VERA project members 2012, IAU Symposium Cosmic Masers - from OH to H0, 287, 391
\reference{} Noriega-Crespo, A., Morris, P., Marleau, F.~R., et al.\ 2004, \apjs, 154, 352 
\reference{} O'Dell, C.~R., \& Wong, K.\ 1996, \aj, 111, 846
\reference{} Olmi, L., \& Cesaroni, R.\ 1999, \aap, 352, 266 
\reference{} Olnon, F.~M.\ 1975, \aap, 39, 217
\reference{} Ossenkopf, V., \& Henning, T.\ 1994, \aap, 291, 943 
\reference{} Osterbrock, D.~E.\ 1989, Research supported by the University of California, John Simon Guggenheim Memorial Foundation, University of Minnesota, et al.~Mill Valley, CA, University Science Books, 1989, 422 p.\
\reference{} Panagia, N., \& Felli, M.\ 1975, \aap, 39, 1 
\reference{} Panagia, N.\ 1973, \aj, 78, 929 
\reference{} Pratap, P., Shute, P.~A., Keane, T.~C., Battersby, C., \& Sterling, S.\ 2008, \aj, 135, 1718 
\reference{} Qiu, K., Zhang, Q., Megeath, S.~T., et al.\ 2008, \apj, 685, 1005
\reference{} Raga, A.~C., Binette, L., Cant\'o, J., \& Calvet, N.\ 1990, \apj, 364, 601
\reference{} Reid, M.~J., Argon, A.~L., Masson, C.~R., Menten, K.~M., \& Moran, J.~M.\ 1995, \apj, 443, 238
\reference{} Riffel, R. A., \& L\"udke, E. 2010, MNRAS, 404, 1449
\reference{} Roberts, H., \& Millar, T.~J. 2007, \aap, 471, 849 
\reference{} Rodr{\'{\i}}guez, L.~F., Gonz{\'a}lez, R.~F., Montes, G., et al.\ 2012, \apj, 755, 152
\reference{} Rodr{\'{\i}}guez, L.~F., Moran, J.~M., Franco-Hern{\'a}ndez, R., et al.\ 2008, \aj, 135, 2370 
\reference{} Rodr{\'{\i}}guez, L.~F., Garay, G., Brooks, K.~J., \& Mardones, D.\ 2005, \apj, 626, 953
\reference{} Rodr{\'{\i}}guez, L.~F., Garay, G., Curiel, S., et al.\ 1994, \apjl, 430, L65 
\reference{} Rodr{\'{\i}}guez, L.~F., Marti, J., Canto, J., Moran, J.~M., \& Curiel, S.\ 1993, RMxAA, 25, 23 
\reference{} Rodr{\'{\i}}guez, L.~F., Curiel, S., Moran, J.~M., Mirabel, I.~F., Roth, M., \& Garay, G.\ 1989, \apjl, 346, L85
\reference{} Rohlfs, K., \& Wilson, T.~L.\ 2004, Tools of radio astronomy, 4th rev.~and enl.~ed., by K.~Rohlfs and T.L.~Wilson.~ Berlin: Springer, 2004,  
\reference{} Rygl, K.~L.~J., Brunthaler, A., Sanna, A., et al.\ 2012, \aap, 539, A79
\reference{} S{\'a}nchez-Monge, {\'A}.\  2011, PhD Thesis. Universitat de Barcelona.
\reference{} S{\'a}nchez-Monge, {\'A}., Palau, A., Estalella, R., Beltr{\'a}n, M.~T., \& Girart, J.~M.\ 2008, \aap, 485, 497 
\reference{} Shchekinov, Y.~A., \& Sobolev, A.~M.\ 2004, \aap, 418, 1045 
\reference{} Shepherd, D.~S., Churchwell, E., \& Wilner, D.~J.\ 1997, \apj, 482, 355 
\reference{} Shirley, Y.~L., Evans, N.~J., II, Young, K.~E., Knez, C., \& Jaffe, D.~T.\ 2003, \apjs, 149, 375 
\reference{} Simpson, J.~P., Cotera, A.~S., Burton, M.~G., et al.\ 2012, \mnras, 419, 211
\reference{} Smith, H.~A., Hora, J.~L., Marengo, M., \& Pipher, J.~L.\ 2006, \apj, 645, 1264
\reference{} Stahler, S.~W., \& Palla, F.\ 2005, The Formation of Stars, by Steven W.~Stahler, Francesco Palla, pp.~865.~ISBN 3-527-40559-3.~Wiley-VCH , January 2005.,  
\reference{} Tafoya, D., G{\'o}mez, Y., \& Rodr{\'{\i}}guez, L.~F.\ 2004, \apj, 610, 827 
\reference{} Takami, M., Chen, H.-H., Karr, J.~L., et al.\ 2012, \apj, 748, 8
\reference{} Takami, M., Karr, J.~L., Koh, H., Chen, H.-H., \& Lee, H.-T.\ 2010, \apj, 720, 155
\reference{} Varricatt, W.~P.\ 2011, \aap, 527, A97
\reference{} Wood, D.~O.~S., \& Churchwell, E.\ 1989, \apjs, 69, 831 
\reference{} Zapata, L.~A., Rodr{\'{\i}}guez, L.~F., Ho, P.~T.~P., Beuther, H., \& Zhang, Q.\ 2006, \aj, 131, 939 
\reference{} Zapata, L.~A., Rodr{\'{\i}}guez, L.~F., Kurtz, S.~E., \& O'Dell, C.~R.\ 2004, \aj, 127, 2252

\end{references}
\end{document}